\documentstyle[times,pramana,epsfig,floats]{ias}
\begin{document}
\mark{{Lattice QCD equation of state}{R.\ V.\ Gavai, S.\ Gupta and S.\ Mukherjee}}
%
%
\def \beq{\begin{equation}}
\def \eeq{\end{equation}}
\def \beqa{\begin{eqnarray}}
\def \eeqa{\end{eqnarray}}
%
%
\def \Ds{D_s}
\def \Dt{D_{\tau}}
\def \real{{\rm Re}\,}
\def \tr{{\rm Tr}\,}
\def \Z{{\cal Z}}
\def \C{{\cal C}}
\def \cs{C_s}
\def \cv{C_{\scriptscriptstyle V}}
\def \pt{p_{\scriptscriptstyle T}}
\def \varss{\sigma_{ss}}
\def \varst{\sigma_{s\tau}}
\def \vartt{\sigma_{\tau\tau}}
\def \lambdams{\Lambda_{\overline{\scriptscriptstyle MS}}}
%
%
\def \etc{{\rm etc.\ }}
\def \eg{{\rm e.g.\ }}
\def \ie{{\rm i.e.\ }}
\def \viz{{\rm viz.\ }}
\def \etal{{\rm et al.}}
\def \jhep{{\sl J.\ H.\ E.\ P.\ }}
\def \np{{\sl Nucl.\ Phys.\ }}
\def \pl{{\sl Phys.\ Lett.\ }}
\def \pr{{\sl Phys.\ Rev.\ }}
\def \prl{{\sl Phys.\ Rev.\ Lett.\ }}

\title{Lattice QCD equation of state : improving the differential method}

\author{ Rajiv\ V.\ Gavai \footnote{E-mail: gavai@tifr.res.in}, 
Sourendu Gupta \footnote{E-mail: sgupta@tifr.res.in} and
Swagato Mukherjee \footnote{Present address: Fakult\"{a}t f\"{u}r Physik,
Universit\"{a}t Bielefeld, D-33615 Bielefeld, Germany. E-mail:
smukher@physik.uni-bielefeld.de} }

\address{ Department of Theoretical Physics, Tata Institute of Fundamental Research,
\\ Homi Bhabha Road, Mumbai 400005, India. }


\abstract{ 
We propose an improvement of the differential method for the computation of the
equation of state of QCD from lattice simulations. In contrast to the earlier
differential method our technique yields positive pressure for all
temperatures including in the transition region. Employing it on temporal
lattices of 8, 10 and 12 sites and by extrapolating to zero lattice spacing
we obtained the pressure, energy density, entropy density, specific heat 
and speed of sound in quenched QCD for $0.9\le T/T_c\le3$.
A comparison of our results is made with those from the dimensional reduction 
approach and a conformal symmetric theory at high-temperature.  
}
\maketitle

\section{Introduction} \label{sc.introduction}

There is growing acceptance of the view that in the ongoing experiments in
Relativistic Heavy Ion Collider (RHIC) at Brookhaven a new form of matter has
been created \cite{rhic}. This new form of matter is thought to be a fluid of
strongly interacting quarks and gluons. In lattice studies of quenched QCD it
was found earlier that the entropy density $s$ \cite{boyd,swagato} and the mean
free time $\tau$, derived from the electrical conductivity \cite{gupta1},
together gave rise to a dimensionless number $\tau
s^{1/3}\approx0.8$ \cite{iitk}. In the non-relativistic limit this dimensionless
number measures the mean free path in units of interparticle spacing, and is
therefore large in a gas but of order unity in a liquid. This 
indicated that the deviation of the energy density ($\epsilon$) and pressure
($P$) in the high temperature phase of QCD from their ideal gas values may be
due to a previously underappreciated feature of the plasma phase--- that it is
far from being a weakly interacting gas.

Earlier expectations that a weakly interacting gas of quarks and gluons would
be formed in the experiments were based on perturbative calculations
\cite{pert} which failed to reproduce these lattice results \cite{boyd}. There
have been many suggestions for the physics implied by the lattice data--- the
inclusion of various quasi-particles \cite{peshier}, the necessity of large
resummations \cite{bir}, and effective models \cite{pisarski} being a few.
Investigation of screening masses also gave evidence for strong departure from
perturbative results \cite{grossman,kari,saumen,owe,saumen2}. Interestingly,
there has been a suggestion that conformal field theory comes closer to the
lattice result \cite{gubser}. This assumes more significance in view of the fact
that a bound on the ratio of the shear viscosity and the entropy density, $s$,
conjectured from the AdS/CFT correspondence \cite{son} lies close to that
inferred from analysis of RHIC data \cite{teaney} and its direct lattice
measurement \cite{nakamura,meyer} as well as the lattice results of a different
transport coefficient \cite{gupta1}.

The equation of state (EOS) is one of the most basic inputs into the analysis
of experimental data. Two decades ago, a method was devised to compute the EOS
of QCD on the lattice\cite{engels}.   However, soon it was found \cite{deng}
that this method yielded negative $P$ near the critical temperature, $T_c$. At
that time it was thought that this problem of the ``differential method'', as
it is called now, is solely due to the use of perturbative formulae for various
derivatives of the coupling. To cure this problem of negative pressure the
non-perturbative ``integral method'' was introduced \cite{engels1,boyd}. It
bypasses the use of perturbative couplings by employing the thermodynamic
relation $F=-PV$ and using a non-perturbative but phenomenologically fitted QCD
$\beta$-function.  If the EOS were to be evaluated by the integral method then
fluctuation measures (\eg the specific heat at constant volume $\cv$) can only
be evaluated through numerical differentiation, which is prone to large errors
\cite{numrec}.   Moreover, the relation $F=-PV$ assumes the system to be
homogeneous. Since the pure gauge phase transition in QCD is of first order the
system is not homogeneous at $T_c$.  Thus one makes an unknown systematic error
in the integral method computation by integrating through $T_c$. This is in
addition to a small systematic error due to setting $P=0$ just below $T_c$ and
the numerical integration errors. Clearly, our confidence in the lattice
results on the EOS would be boosted if an entirely different method of EOS
determination yields the same results: it would tantamount to a good control
over many systematic errors in both.  

In this paper we propose a modification of the differential method which gives
positive pressure over the entire temperature range for even relatively coarser
lattices . We choose the temporal lattice spacing ($a_\tau$) to set the scale
of the theory, in contrast to the choice of the spatial lattice spacing ($a_s$)
in the approach of \cite{engels}.  This change of scale is analogous to the use
of different renormalization schemes.  As a consequence, our method could be
called the {\sl t-favoured scheme} and the method of Ref.\ \cite{engels} may
be called the {\sl s-favoured scheme}.  In fact, in a different context, 
this choice of scale has already been used in Ref.\ \cite{ejiri}.  Here
we show that this choice leads to  positive pressure for the entire
temperature range, even when one uses one-loop order perturbative couplings.
Since the operator expressions are derived with an asymmetry between the two
lattice spacings $a_s$ and $a_\tau$, the s-favoured and t-favoured schemes give
different expressions for the pressure. In that sense the use of t-favoured
scheme is tantamount to the use of better operators.

Being a differential method the t-favoured scheme can be easily extended for
the calculation of fluctuation measures like $\cv$, following the formalism developed
in Ref.\ \cite{swagato}. In a theory with only gluons there is only this one
fluctuation measure. Related to this is a kinetic variable, the speed of sound,
$\cs$, which can also be evaluated in any operator method. We report measurements
of both in the temperature range $0.9T_c \le T \le 3T_c$ through a continuum
extrapolation of results obtained using successively finer lattices.

Not only do these quantities provide further tests of all the models which try
to explain the lattice data on the EOS they also have direct physical
relevance to experiments at RHIC.  In a canonical ensemble the specific heat at
constant volume is a measure of energy fluctuations.  It was suggested in Ref.\ 
\cite{stodolsky} that event-by-event temperature fluctuation in the heavy-ion
collision experiments can be used to measure $\cv$.  The speed of sound, on the
other hand, controls the expansion rate of the fire-ball produced in the
heavy-ion collisions. Thus the value of $\cs$ is an important parameter in the
hydrodynamic studies. It has been noted that the magnitude of elliptic flow in
heavy-ion collisions is sensitive to the value of $\cs$ \cite{blaizot}.

The measurement of $\cv$ and $\cs$ also directly test the relevance of
conformal symmetry to finite temperature QCD. QCD is known to generate the
scale, $\Lambda_{QCD}$, dynamically and thus break conformal invariance.
The strength of the breaking of this symmetry at any scale is parametrized by
the $\beta$-function.  An effective theory which
reproduces the results of thermal QCD at long-distance scales could still be
close to a conformal theory. The result of Ref.\ \cite{gubser} for the entropy
density, $s$, in a Yang-Mills theory with four supersymmetry charges (${\cal
N}=4$ SYM) and large number of colours, $N_c$, at strong coupling, is
\beqa 
  \frac s{s_0} &=& f(g^2 N_c), \quad{\rm where}\quad 
  s_0 = \frac23 \pi^2 N_c^2T^3 \quad{\rm and}\quad \nonumber\\
  f(x)&=&\frac34+\frac{45}{32}\zeta(3)x^{-3/2}+\cdots ,
\label{eq.sym}
\eeqa
$g$ being the Yang-Mills coupling \footnote{We thank Igor Klebanov for
pointing out that the factor $x^{-3/2}$ in the right hand side of Eq.
(\ref{eq.sym}) appears in early literature as $(2x)^{-3/2}$ due to a different
normalization of $g^2N_c$. }.  For our case of $N_c=3$, the well-known result
for the ideal gas, $s_0=4(N_c^2-1)\pi^2T^3/45$ takes into account, through the
factor $N_c^2-1$, the relatively important difference between a $SU(N_c)$ and
an $U(N_c)$ theory.

The paper is organized as follows. In the next section we present the formalism
and lead up to the measurement of $\cv$ and $\cs^2$ on the lattice in Section
2.2. In Section 3 we give details of our simulations
and our results. Finally, in Section 4 we present a
discussion of the results.

\section{Formalism} \label{sc.formalism}

Various derivatives of the partition function, $\Z(V,T)$, where $V$ is the
volume and $T$ the temperature, lead to thermodynamic quantities of interest. In
particular the energy density, $\epsilon$, and the pressure, $P$, are given by
the first derivatives of $\ln\Z$,
\beq
   \epsilon = \left(\frac TV\right)
     \left.T\frac{\partial\ln\Z(V,T)}{\partial T}\right|_V
   \qquad{\rm and}\qquad
   P = \left(\frac TV\right)\left.V\frac{\partial\ln\Z(V,T)}{\partial V}\right|_T.
\label{eq.energy-pressure}
\eeq
The second derivatives are measures of fluctuations.  In the absence of
chemical potentials a change of volume of a relativistic gas alters its
pressure by changing particle numbers. As a result there is only one second
derivative, namely the specific heat at constant volume--- 
\beq
   \cv = \left.\frac{\partial\epsilon}{\partial T}\right|_V.
\label{eq.specht}\eeq

Using thermodynamic identities, the expression for the speed of sound 
can be recast in the form 
\beq
   \cs^2 \equiv \left.\frac{\partial P}{\partial\epsilon}\right|_s
     = \left.\frac{\partial P}{\partial T}\right|_V
      \left(\left.\frac{\partial\epsilon}{\partial T}\right|_V\right)^{-1}
     = \frac{s/T^3}{\cv/T^3},
\label{eq.sound}\eeq
where we have used the thermodynamic identity
\beq
   \left.\frac{\partial P}{\partial T}\right|_V = 
      \left.\frac{\partial S}{\partial V}\right|_T 
   \quad{\rm and}\quad
      \left.\frac{\partial S}{\partial V}\right|_T =
    s = \frac{\epsilon+P}T,
\label{eq.entropy}\eeq
in conjunction with the definition of the total entropy, $S$, and the 
entropy density, $s$, above. Note that all these relations are valid
for full QCD with dynamical quarks (without quark chemical potentials)
as well as in the quenched approximation which this work deals with 
exclusively. 

A caveat about the first equality in Eq.\ (\ref{eq.sound}) is in
order. This remarkable formula (a generalization of a result first
obtained in 1687 by Newton) equating a kinetic quantity, $\cs^2$,
to a thermodynamic derivative is true for a homogeneous system. For
a phase mixture at a first order phase transition there are kinetic
processes, such as condensation of a fog, which cause this formula
to break down \cite{landau}. The lore that $\cs^2=0$ at $T_c$ is
due to the overly naive argument that $P$ remains continuous while
$\epsilon$ undergoes a discontinuous change. In fact, the best that
thermodynamics can do is to evaluate this formula in a limiting
sense as one approaches $T_c$ either from above or below. The values
of $\cs$ in these two limits need not even be continuous at a first
order transition \cite{gock}.

\subsection{Energy density and pressure} \label{sc.eos}

In order to distinguish between $T$ and $V$ derivatives, the differential method
formulate the theory on a $d+1$ dimensional asymmetric lattice having
different lattice spacings in the spatial ($a_s$) and the temporal
($a_{\tau}$) directions. If the number of lattice sites in the two directions
are $N_s$ and $N_\tau$, then $T=(N_\tau a_\tau)^{-1}$ is the temperature and
$V=(N_s a_s)^d$ is the volume of the system. The derivatives needed for the
thermodynamics are---
\beq
   T\left.\frac{\partial}{\partial T}\right|_V = 
       -a_\tau\left.\frac{\partial}{\partial a_\tau}\right|_{a_s}
   \qquad{\rm and}\qquad
   V\left.\frac{\partial}{\partial V}\right|_T = 
       \frac{a_s}d\left.\frac{\partial}{\partial a_s}\right|_{a_\tau},
\label{allders}\eeq
holding $N_s$ and $N_\tau$ fixed.

In the t-favoured scheme we
introduce the anisotropy parameter $\xi$ and the scale $a$ by the relations,
\beq 
  \xi=\frac{a_s}{a_{\tau}}, \qquad{\rm and}\qquad a=a_{\tau}.  
\label{tfavoured}\eeq  
The partial derivatives with respect to $T$ and $V$ can then be written 
in terms of these new variables as
\beq
   T\left.\frac{\partial}{\partial T}\right|_{V} =
       \xi\left.\frac{\partial}{\partial \xi}\right|_{a} -
       a\left.\frac{\partial}{\partial a}\right|_{\xi},
   \qquad{\rm and}\qquad
   V\left.\frac{\partial}{\partial V}\right|_{T} =
       \frac\xi d\left.\frac{\partial}{\partial \xi}\right|_{a}.
\label{eq.lat-der}\eeq
One obtains the second expression by writing $a_s=a\xi$ and taking
a partial derivative keeping $a$ fixed. For the first expression,
one takes a derivative with respect to $a$ and then introduces constraints
on the differentials $d\xi$ and $da$ in order to keep $a_s$ fixed.
This choice of scale $a=a_\tau$ seems to be natural, since
most numerical work at finite temperature sets the scale by $T=1/N_\tau
a_\tau$.  For example, continuum limits are taken at fixed physics
by keeping $T$ fixed while changing $N_\tau$ and $a_\tau$ simultaneously.
This is done not only when symmetric lattices are used, but also
when the simulation is performed with asymmetric lattices \cite{aniso}.

In the s-favoured method \cite{engels}, by contrast, the scale of the
theory is set by the spatial lattice spacing, $a=a_s$, at every $\xi$ and
only after taking the $\xi\to1$ limit does the natural choice of scale
emerge. The corresponding derivatives in this case are
\beq
   T\left.\frac{\partial}{\partial T}\right|_{V} =
       \xi\left.\frac{\partial}{\partial \xi}\right|_{a}
   \qquad{\rm and}\qquad
   V\left.\frac{\partial}{\partial V}\right|_{T} =
       \frac\xi d\left.\frac{\partial}{\partial \xi}\right|_{a}
       +\frac a d\left.\frac{\partial}{\partial a}\right|_\xi.
\label{eq.sfav-lat-der}\eeq

On the anisotropic lattice the partition function of a pure
gauge $SU(N_c)$ theory with the Wilson action is defined as  
\beqa
   {\cal Z}(V,T)&=&\int{\cal{D}}U e^{-S[U]},
   \qquad{\rm where}\qquad \nonumber \\
   S[U]&=&K_s\sum_{x,ij=1}^dP_{ij}(x)+K_\tau\sum_{x,i=1}^dP_{0i}(x).
\label{eq.part-func}
\eeqa
Periodic boundary conditions are imposed in all directions. The
plaquette variables are $P_{\alpha\beta}(x)=1-\real ~{\rm tr}
U_{\alpha\beta}(x)$, $U_{\alpha\beta}(x)$ is the ordered product of link
matrices taken anticlockwise around the plaquette, starting at the site
$x$ and in the plane specified by the directions $\alpha$ and
$\beta$. We introduce the notation for the average plaquettes
$P_s=2\sum P_{ij}(x)/ d(d-1)N_s^dN_\tau$ and $P_\tau=\sum P_{0i}(x)
/dN_s^dN_\tau$. Since the plaquette operators have no explicit
dependence on $a$ and $\xi$ the derivatives with respect to these
quantities vanish. The couplings may be written as  
\beq 
   K_s=\frac{2N_c}{\xi g_s^2},
   \qquad{\rm and} \qquad 
   K_\tau=\frac{2N_c\xi}{g_\tau^2},
\label{eq.coupling}\eeq
leading to
\beq
   \xi\frac{\partial K_s}{\partial\xi} = 
       -K_s+2N_c\frac{\partial g_s^{-2}}{\partial\xi},
   \qquad{\rm and}\qquad
   \xi\frac{\partial K_\tau}{\partial\xi} = 
        K_\tau + 2N_c\xi^2 \frac{\partial g_\tau^{-2}}{\partial\xi}.
\label{eq.coup-der-1}\eeq

Next, using the derivatives in Eq.\ (\ref{eq.lat-der}) along with
the definitions of $P$ and $\epsilon$ (see Eq.\ \ref{eq.energy-pressure}) one
obtains, from the partition function of Eq.\ (\ref{eq.part-func}), the
expressions
\beqa
   a^{d+1}\epsilon &=& -\frac{d}{\xi^d} \left[ \frac{d-1}2 \xi K_s'\Ds +
      \xi K_\tau' \Dt \right] +
       \frac{d}{\xi^d} \left[ \frac{d-1}2\,a\frac{\partial K_s}{\partial a}\Ds
           +a\frac{\partial K_{\tau}}{\partial a}\Dt \right]
   \quad{\rm and}
   \nonumber \\ 
   a^{d+1}P &=& -\frac1{\xi^d} \left[ \frac{d-1}2 \xi K_s'\Ds 
       +  \xi K_\tau'\Dt \right].
\label{eq.energy-pressure-1}\eeqa
where primes denote derivative with respect to $\xi$. In order to remove the
trivial ultraviolet divergence in these quantities, present even in the free
case, a subtraction of the corresponding $T=0$ values is made, yielding
$D_i=\langle P_i\rangle -\langle P_0\rangle$ above.   Here $P_0=2\sum
P_{\alpha\beta}(x)/d(d+1)N_s^dN_\tau$ is the average plaquette value at $T=0$,
evaluated with periodic boundary conditions in all directions and with very
large $N_\tau=N_s$.

To determine the couplings $K_i'$ we use the weak coupling definitions
\cite{hasenfratz}
\beq 
   \frac1{g_i^2(a,\xi)}=\frac1{g^2(a)}+c_i(\xi)+O\left[g^2(a)\right]
   \quad(i=s,\tau).
\label{eq.wcoupling}\eeq
With the condition that $c_i(\xi=1)=0$, this is actually an expansion of
the anisotropic lattice couplings $g_i(a,\xi)$ around the isotropic
lattice coupling $g(a)$. 
With the usual definition, $\alpha_s=g^2/4\pi$, the $\beta$-function is---
\beq
   B(\alpha_s) = \frac\mu2\,\frac{\partial\alpha_s}{\partial\mu}
   \qquad{\rm giving}\qquad
   a\frac{\partial g^{-2}}{\partial a}=\frac{B(\alpha_s)}{2\pi\alpha_s^2}.
\label{betafn}\eeq
For a $3+1$ dimensional theory one has $B(\alpha_s) =
-(33-2N_f)\alpha_s^2/12\pi +{\cal O}(\alpha_s^3)$. In terms of the
functions $c_s$ and $c_\tau$ introduced in Eq.\ (\ref{eq.wcoupling})
and the $\beta$-function above one can rewrite the derivatives of the
couplings as
\beqa
\nonumber
   a \frac{\partial K_s}{\partial a} &=& 
       \frac{N_cB(\alpha_s)}{\pi\alpha_s^2\xi},
           \qquad{\rm and}\qquad
   \xi\frac{\partial K_s}{\partial\xi} = -K_s+2N_c c_s', \\
   a\frac{\partial K_{\tau}}{\partial a} &=& 
          \frac{N_c\xi B(\alpha_s)}{\pi\alpha_s^2},
           \qquad{\rm and}\qquad
    \xi\frac{\partial K_\tau}{\partial\xi} = 
          K_\tau + 2N_c\xi^2 c_{\tau}'.
\label{eq.coup-der-1a}\eeqa
The quantities $c_s'$ and $c_\tau'$ have been computed to one-loop
order in the weak coupling limit for $SU(N_c)$ gauge theories in 3+1
dimensions \cite{karsch}.

\subsection{The specific heat and speed of sound} \label{sc.cvcs}

It was pointed out in Ref.\ \cite{swagato} that the specific heat can be most easily
obtained by working with the conformal measure, 
\beq
   \C = \frac\Delta\epsilon
   \qquad{\rm and}\qquad
   \Gamma = T\left.\frac{\partial\C}{\partial T}\right|_V,
\label{eq.gamma}\eeq
where $\Delta=\epsilon-3P$.
Then, using Eqs.\ (\ref{eq.sound}, \ref{eq.entropy}, \ref{eq.gamma})
it is straightforward to see that
\beqa
   \frac\cv{T^d} &=& \left(\frac{\epsilon/T^{d+1}}{P/T^{d+1}}\right)
     \left[\frac s{T^d}+\frac\Gamma d \frac\epsilon{T^{d+1}}\right]
   \quad{\rm and}\quad \nonumber\\
   \cs^2 &=& \left(\frac{P/T^{d+1}}{\epsilon/T^{d+1}}\right)
      \left[ 1+\frac{\Gamma\epsilon/T^{d+1}}{ds/T^d}\right]^{-1}.
\label{eq.cvcs}\eeqa

One needs the expression for $\Gamma$ in terms of the plaquettes
in order to proceed. To this end we introduce the two functions
\beqa
   F(\xi, a) &=& \frac{\Delta a^{d+1}\xi^d}d = a
     \left[ \frac{d-1}2\,\frac{\partial K_s}{\partial a}\Ds
       +\frac{\partial K_{\tau}}{\partial a}\Dt \right]
  \qquad {\rm and} \qquad
  \nonumber \\ 
   G(\xi, a) &=& \frac{-\epsilon a^{d+1}\xi^d}d =
      \xi\left[ \frac{d-1}2\,K_s' \Ds + K_{\tau}' \Dt \right] - F(\xi, a).
\label{eq.F-G}\eeqa
Since $\C=-F/G$, one finds that
\beq
   \Gamma = -{\cal C}\frac TF \left.\frac{\partial F}{\partial T}\right|_V
        + {\cal C}\frac TG\left.\frac{\partial G}{\partial T}\right|_V.
\label{eq.gamma-1}\eeq

The derivatives of $F$ and $G$ will involve the variances and covariances of
the plaquettes and the second derivatives of the couplings.
These second derivatives of the couplings are--- 
\beqa
\nonumber
   a\frac{\partial\xi K_s'}{\partial a} =
     -\frac{B(\alpha_s)}{2\pi\alpha_s^2\xi}, &\quad&
   \xi^2K_s''= \frac 2{g_s^2\xi}-2c_s'+\xi c_s'', \\
\nonumber
   a\frac{\partial\xi K_\tau'}{\partial a} =
      \frac{\xi B(\alpha_s)}{2 \pi\alpha_s^2}, &\quad&
   \xi^2K_\tau'' = 2c_\tau'+\xi c_\tau'', \\
   a^2\frac{\partial^2 K_s}{\partial a^2}=
     -\frac{B(\alpha_s)}{2\pi\alpha_s^2\xi} = -a\frac{\partial K_s}{\partial a},
     &\quad&
   a^2\frac{\partial^2 K_\tau}{\partial a^2}=
      -\frac{\xi B(\alpha_s)}{2 \pi \alpha_s^2} = -a\frac{\partial K_\tau}{\partial a}.
\label{eq.coup-der-2}\eeqa
The numerical values of $c_i''$ 's have been evaluated in Ref.\ \cite{swagato}.

Turning now to the derivatives of $F$ and $G$ in Eq.\ (\ref{eq.F-G}) one obtains
\beqa
  \xi\frac{\partial F}{\partial \xi} &=& 
     \xi a \left[ \frac{d-1}2\frac{\partial K_s'}{\partial a}\Ds
     +\frac{\partial K_{\tau}'}{\partial a}\Dt \right] + 
     \xi a \left[ \frac{d-1}2\frac{\partial K_s}{\partial a}\Ds'
     +\frac{\partial K_{\tau}}{\partial a}\Dt' \right],
  \quad {\rm and} \nonumber \\ 
  a\frac{\partial F}{\partial a} &=& 
     a \left[ \frac{d-1}2\frac{\partial K_s}{\partial a}\Ds
     +\frac{\partial K_{\tau}}{\partial a}\Dt \right] + 
     a^2 \left[ \frac{d-1}2\frac{\partial^2 K_s}{\partial a^2}\Ds
     +\frac{\partial^2 K_{\tau}}{\partial a^2}\Dt \right] \nonumber \\
     &+& 
     a^2 \left[ \frac{d-1}2\frac{\partial K_s}{\partial a} \frac{\partial \Ds}{\partial a}
     +\frac{\partial K_{\tau}}{\partial a} \frac{\partial \Dt}{\partial a}
     \right].
\label{eq.ders1}\eeqa
Also from Eq. (\ref{eq.F-G}) it follows
\beqa
 \xi\frac{\partial G}{\partial \xi} &=& 
   \xi \left[ \frac{d-1}{2} K_s' \Ds + K_{\tau}' \Dt \right] + 
   \xi^2 \left[ \frac{d-1}{2} K_s'' \Ds + K_{\tau}'' \Dt \right] \nonumber \\
   &+&\xi^2 \left[ \frac{d-1}{2} K_s' \Ds' + K_{\tau}' \Dt' \right] - 
   \xi\frac{\partial F}{\partial \xi},
 \qquad {\rm and}\nonumber \\
 a\frac{\partial G}{\partial a} &=& 
   \xi a \left[ \frac{d-1}{2} \frac{\partial K_s'}{\partial a}\Ds
   +\frac{\partial K_{\tau}'}{\partial a}\Dt \right] \nonumber \\
   &+&\xi a \left[ \frac{d-1}{2} K_s'\frac{\partial \Ds}{\partial a} + 
    K_{\tau}'\frac{\partial \Dt}{\partial a}\right] -
   a\frac{\partial F}{\partial a}.
\label{eq.ders2}\eeqa
Since the plaquette operators do not explicitly depend on $\xi$ and $a$ one
can easily take the derivatives of the vacuum subtracted plaquette expectation
values. These are
\beqa
 \xi D_i'&=&-dN_\tau N_s^d\left[\frac{d-1}2\xi K_s'\sigma_{si}
      +\xi K_\tau'\sigma_{\tau i}\right],
 \qquad {\rm and} \nonumber \\
 a\,\frac{\partial D_i}{\partial a} &=& -dN_\tau N_s^d
     \left[ \frac{d-1}{2} a\frac{\partial K_s}{\partial a} \sigma_{si} +
     a\frac{\partial K_{\tau}}{\partial a} \sigma_{\tau i}\right],
\label{eq.derexpr}\eeqa
where $\sigma_{i j}=\langle D_i D_j\rangle-\langle D_i\rangle\langle
D_j\rangle$. Throughout this paper we will refer to $\sigma_{ij}$ 
($i\ne j$) as `variances of plaquettes' and $\sigma_{ii}$ as 
`covariances of plaquettes'. Note that Eq.\ (\ref{eq.cvcs}) 
implies that $\cv$ and
$\cs$ should be independent of the volume. Consistent with this,
the derivatives in Eqs.\ (\ref{eq.ders1}, \ref{eq.ders2}) seem to
be non-extensive. However, there is an explicit volume factor,
$N_\tau N_s^d$, in Eq.\ (\ref{eq.derexpr}). The resolution is that
away from a critical point the variances and covariances of the
plaquettes scale as $1/V$, which is a consequence of the central
limit theorem. 

Certainly if each plaquette variable could be considered to be
fluctuating randomly around its mean value then the application of
the central limit theorem would be clear. Before proceeding,
we emphasize that both the plaquette variables defined here are
summed over all spatial orientations, and hence are invariant under
spatial rotations. In the notation of Ref.\ \cite{saumen}, they are
projected on the $A_1^{++}$ channel. Thus, their covariances are
integrals over the $A_1^{++}$ plaquette correlation function. If
plaquette correlations had a finite range, then again these terms
would be linear in volume if $N_s$ were sufficiently large. However, if the
$A_1^{++}$ correlation length associated with plaquettes becomes
infinite, then, in the thermodynamic limit, this term would grow faster
than the remainder. Consistently, at a second order phase transition,
where this is expected, $\cv$, as defined in Eq.\ (\ref{eq.cvcs})
would scale non-trivially with volume according to the critical
exponents of the theory. Such behaviour has been found in the SO(3)
gauge theory \cite{gavai}.

\subsection{Final expressions}

Expressions for the energy density and the pressure in the usual form are obtained from
Eq.\ (\ref{eq.energy-pressure-1}) by multiplying by appropriate powers
of $N_\tau$. In the isotropic ($\xi=1$) limit and for 3+1 dimensions we get 
\beqa
  \frac\epsilon{T^4} &=& 6N_cN_\tau^4 \left[\frac{\Ds-\Dt}{g^2}
     -(c_s'\Ds+c_{\tau}'\Dt) \right] + 
     6N_cN_\tau^4 \frac{B(\alpha_s)}{2 \pi \alpha_s^2} \biggl[\Ds+\Dt\biggr]
   \qquad{\rm and}
   \nonumber \\ 
  \frac{P}{T^4} &=& 2N_cN_\tau^4 \left[\frac{\Ds-\Dt}{g^2}
     -(c_s'\Ds+c_{\tau}'\Dt) \right].
\label{eandp}\eeqa
On comparing these expressions with those obtained using the s-favoured
scheme \cite{engels}, one can easily see that the new
expression for pressure is exactly $1/3$ of the old expression of
the energy density. Since the energy density in the s-favoured scheme
comes out to be non-negative at all temperatures and on all temporal sizes
$N_\tau$ , our new expression for the
pressure is therefore expected to give non-negative pressure always.
The expression for the interaction measure
\beq
 \frac{\Delta}{T^4}=\frac{(\epsilon-3P)}{T^4}=
     6N_cN_\tau^4 \frac{B(\alpha_s)}{2 \pi \alpha_s^2}\biggl[\Ds+\Dt\biggr],
\label{delta}\eeq
is same, and also positive, for both the cases. Since both the pressure and the
interaction measure are non-negative in the t-favoured operator formalism, the
energy density must also be non-negative.

Note that $\Delta$ contains $B(\alpha_s)$ as a factor, but this
explicit breaking of conformal symmetry may be compensated by the
vanishing of the factor $\Ds+\Dt$. To determine the coupling $g^2$,
throughout this work, we use the method suggested in Ref.\ \cite{gupta},
where the one-loop order renormalized couplings have been evaluated
by using $V$-scheme \cite{lepage} and taking care of the scaling
violations due to finite lattice spacing errors using the method
in Ref.\ \cite{edwards}.

The expressions for $\xi$ and $a$ derivatives of $F(\xi,a)$ in Eq.\
(\ref{eq.ders1}) can be combined by using the form of the lattice derivatives
in Eq.\ (\ref{eq.lat-der}) to get the temperature derivative of $F(\xi,a)$.
Finally inserting the derivatives of the coupling (see Eq.\
\ref{eq.coup-der-1} and Eq.\ \ref{eq.coup-der-2}), taking the $\xi \to 1$
limit, and specializing to $d=3$ we get--- 
\beqa
\nonumber
 T\left.\frac{\partial F}{\partial T}\right|_V &=&
  \frac{B(\alpha_s)}{2\pi\alpha_s^2}\left[ \Dt-\Ds \right]
   +6N_c N_{\tau} N_s^3 \biggl[\frac{B(\alpha_s)}{2\pi\alpha_s^2}\biggr]^2
      \left[ \varss+\vartt+2\varst \right] \\
   &-&6N_c N_{\tau} N_s^3 \frac{B(\alpha_s)}{2\pi\alpha_s^2}
      \left[ \frac{\vartt-\varss}{g^2}+
      c_s'\varss+c_{\tau}'\vartt+(c_s'+c_{\tau}')\varst \right] . \quad
\label{eq.dfdt} \eeqa
Proceeding in the similar way as before, in the $\xi \to 1$ limit in
$d=3$, one obtains---
\beqa
\nonumber
 T\left.\frac{\partial G}{\partial T}\right|_V &=&
  \frac{\Ds+\Dt}{g^2}- c_s'\Ds + 3 c_{\tau}'\Dt + c_s''\Ds + c_{\tau}''\Dt
  - \frac{B(\alpha_s)}{2\pi\alpha_s^2}\left[ \Dt-\Ds \right] \\
  \nonumber
  &-& 6N_cN_{\tau}N_s^3
   \left[\frac{\sigma_{s,s}+\sigma_{\tau,\tau}-2\sigma_{s,\tau}}{g^4}
    +\frac{2(c_{\tau}'\sigma_{\tau,\tau}+c_s'\sigma_{s,\tau}-c_s'\sigma_{s,s}
    -c_{\tau}'\sigma_{s,\tau})}{g^2} \right] \\
    \nonumber
    &+& 6N_cN_{\tau}N_s^3 \left[ {c_s'}^2\sigma_{s,s}+
    {c_\tau'}^2\sigma_{\tau,\tau}+2c_s'c_\tau'\sigma_{s,\tau} \right]
    - T\left.\frac{\partial F}{\partial T}\right|_V \\
   &+& 6 N_c N_{\tau} N_s^3 \frac{B(\alpha_s)}{2\pi\alpha_s^2}
   \left[ \frac{\vartt-\varss}{g^2} + c_s'\varss+c_{\tau}'\vartt+
    (c_s'+c_{\tau}')\varst \right] .
\label{eq.dgdt} \eeqa

For $g \to 0$, \ie, in the weak-coupling limit, the dominant contribution to
all the plaquettes is of order $g^2$ \cite{heller}.  Hence, in this limit,
$D_i \propto g^2$, and $\Delta/T^4\propto g^2$. In the weak-coupling limit,
therefore, $\Delta/T^d$ can be neglected in comparison with $\epsilon/T^d$.
The scaling of $D_i$ also implies that $\sigma_{ij}\propto g^4$, as a result
of which $F$ and its temperature derivative are negligible in this limit
compared to $G$ and its derivative. Consequently, $\Gamma\to0$ in this limit,
resulting in $\cv/T^d \to (d+1)\epsilon/T^{d+1}$ and $\cs^2\to1/d$.  Note that
in any conformal invariant theory in $d+1$ dimensions one has $\epsilon=dP$,
\ie, $\C=\Gamma=0$, and hence, by Eq.\ (\ref{eq.cvcs}), one has identical
results--- $\cs^2=1/d$ and $\cv/T^d=(d+1)\epsilon/T^{d+1}$.

\subsection{On the method} \label{sc.method}

While the expressions in Eq.\ (\ref{eandp}) look different from those in Ref.\
\cite{engels}, one may argue \cite{ack} that standard formul\ae{} for 
change of variables (from the set $\{\xi,a_\tau\}$ to $\{\xi,a_s\}$) can be
used to show that both the expressions are identical. However, this conclusion
follows only if one also demands the values of the couplings $g_s^2$ and 
$g_\tau^2$ to remain the same under the change of the scale from $a_s$ to 
$a_\tau$.  As we argue below, this is not true when the weak coupling 
expressions [ Eq.\ (\ref{eq.wcoupling})] are used for the couplings.

As can be seen form Eq.\ (\ref{eq.wcoupling}) the Karsch coefficients
$c_i(\xi)$'s are differences between the isotropic and anisotropic couplings.
Hence they do not depend on the scale $a$ of the isotropic lattice, but only
on the parameter which quantifies the difference between the isotropic and
the anisotropic lattice, \ie, the anisotropy parameter $\xi$. Thus a change of
scale from $a_s$ to $a_\tau$ does not change these Karsch coefficients. In
Appendix \ref{app.a} we prove this explicitly. Given that the Karsch coefficients
are same for both the t-favoured and the s-favoured schemes, from Eq.\
[\ref{eq.wcoupling}] it follows that the anisotropic coupling constants
$g_i(a,\xi)$ are different for the two schemes due to the scale dependence of
the isotropic coupling constant $g(a)$.  Therefore the expressions for
$\epsilon$ and $P$ are different at finite (but small) lattice spacing in the
two different approaches.  Since the s-favoured and t-favoured schemes are
different due to the scale dependence of the isotropic coupling constant
$g(a)$, the difference between the expressions in both the schemes goes as $\ln
a$, compared to the $1/a^2$ cut-off dependence of the lattice Wilson action.
Hence, the difference between the two methods is tantamount to modifying the
operators. Moreover, for the usual choice of scale setting by $T=1/N_\tau
a_\tau$, our approach corresponds to the natural choice of scale in Eq.\
[\ref{eq.wcoupling}].  It is expected that the results from
both the methods will match for very large temporal lattice size $N_\tau$. 
However, as is true with the improvement program in general, on small lattices
the better operators--- t-favoured method in this case --- should lead
to results with lesser artifact errors or alternatively positive pressure
at even $T \le T_c$.

\begin{figure}[t!]
\begin{center}
\includegraphics[scale=0.7]{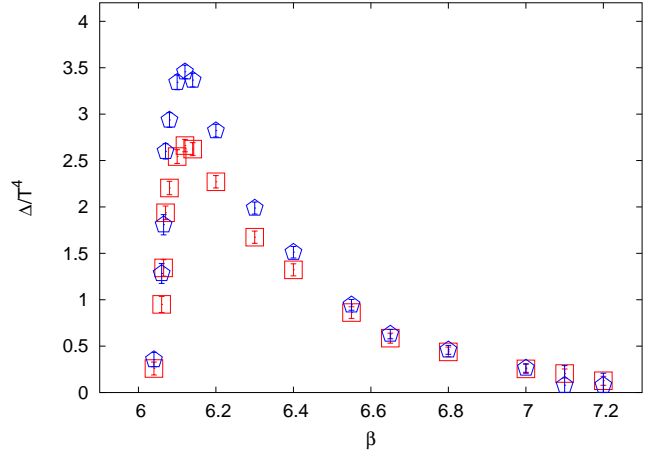}
\end{center}
\caption{$\Delta/T^4$ as a function of the bare coupling $\beta$ using a
non-perturbative (squares) and one-loop order perturbative (pentagons)
$\beta$-function, $B(\alpha_s)$. The results agree for $\beta\ge6.55$. The plaquette
values for $N_\tau=8$ and the values of the non-perturbative $\beta$-function are
taken from Ref.\ [2].}
\label{fig.delta-tfav-int}
\end{figure}

While the t-favoured method improves the differential method, leading to positive
pressure, it still requires the use of perturbative couplings.   On the other hand,
the integral method evades them but at the cost of the assumption of homogeneity. For
small volumes used in actual simulations, one may feel reassured by its test in form
of agreement of results with other methods such as the differential method. Note that
the expression for $\Delta/T^4$ is identical for both the integral method and the
t-favoured scheme. It depends on the $\beta$-function, $B(\alpha_s)$ in Eq.\
[\ref{betafn}]. A non-perturbatively determined $\beta$-function permits the integral
method to lead to fully non-perturbative EOS.  However, one usually fits a
phenomenological ansatz to extract it from a range of couplings $6/g^2$ with their
associated systematic uncertainties.  The differential method could also employ such
a $\beta$-function but for internal consistency we require that the Karsch
coefficients and $B(\alpha_s)$ be obtained at the same order, \ie at one-loop order
in the present state of the art. 

The two methods must agree if one uses sufficiently small lattice spacings,
\viz when the use of perturbative couplings is justified in the differential
method computation and on large enough volumes.  A a comparison between the
values of $\Delta/T^4$ extracted for a given $N_\tau$ using the two approaches
would reveal at what $T$ the two methods become close to each other. Using
asymptotic scaling, one could also then find the minimum value of $N_\tau$ 
required for the same level of agreement as a function of $T$. Such a
comparison is shown in Figure\ \ref{fig.delta-tfav-int}, which demonstrates
that a bare coupling of $\beta\ge6.55$ should suffice to give an agreement
between the t-favoured scheme and the integral method. For $\beta\le6.55$ use
of one-loop order perturbative Karsch coefficients may give rise to some
systematic effects. A comparison with the non-perturbatively determined Karsch
coefficients \cite{ejiri,engels2} shows that the difference between the
perturbative and non-perturbative values are significant. For example, while at
around $\beta=6.55$ the one-loop order perturbative and non-perturbative $c_i'$
differ by $\sim20\%$, around $\beta=6$ this difference increases to $\sim80\%$.

In the present work we show that within the framework of differential method it
possible to get a positive pressure for all temperatures if one uses the
better operators of the t-favoured scheme. This is so in spite of the use of
one-loop order perturbative Karsch coefficients. However, the use of one-loop
order perturbative Karsch coefficients \cite{karsch,swagato} may give some
systematic effects if the lattice spacing is not small enough.

\section{Simulations and results} \label{sc.results}

Our simulations have been performed using the Cabbibo-Marinari
pseudo-heatbath algorithm with Kennedy-Pendleton updating of three
$SU(2)$ subgroups on each sweep. Plaquettes were measured on each
sweep. For each simulation we discarded around 5000 initial sweeps
for thermalization. We found that the maximum value for the integrated
autocorrelation time for the plaquettes is about 12 sweeps for the
$T=0$ run at $\beta=6$ and the minimum was 3 sweeps for the $T=3T_c$
run for $N_\tau=12$. Table \ref{tb.simulation} lists the details
of these runs.  All errors were calculated by the jack-knife method,
where the length of each deleted block was chosen to be at least
six times the maximum integrated autocorrelation time of all the
simulations used for that calculation.

In Ref.\ \cite{saumen1} it was shown that, at sufficiently high temperature,
finite size effects are under control if one chooses $N_s=(T/T_c)N_\tau
+2$ for the asymmetric ($N_{\tau} \times N_s^3$) lattice. We have
chosen the sizes of the lattices used at finite $T$ based on this
investigation. Close to $T_c$ the most stringent constraint on
allowed lattice sizes comes from the $A_1^{++}$ screening mass
determined in Ref.\ \cite{saumen2}. Among the temperature values we
investigated, this screening mass is smallest at $1.25T_c$ where
it is a little more than $2T$. The choice of $N_s=2N_\tau+2$ satisfies
this constraint sufficiently. If future work pushes closer to $T_c$,
then larger values of $N_s$ need to be used in view of the further
decrease in the $A_1^{++}$ screening mass. At $T=0$ the constraints
are simpler because glueball masses are larger, and also smoother
functions of $\beta$. For the symmetric ($N_s^4$) lattices we have
chosen $N_s = 22$ as the minimum lattice size and scaled this up
with changes in the lattice spacing in accordance with the analysis
done in Ref.\ \cite{swagato}.

\begin{table}[t!]
\caption{The coupling ($\beta$), lattice sizes ($N_\tau\times N_s^3$), statistics and
symmetric lattice sizes ($N_s^4$) are given for each temperature. Statistics means
number of sweeps used for measurement of plaquettes after discarding for
thermalization.}
 \begin{tabular}{|c|c|c c|c c|} \hline
    $T/T_c$&$\beta$&&Asymmetric Lattice&&Symmetric Lattice \\
    &&size&stat.&size&stat. \\ \hline

    &6.0000&$8 \times 18^3$&1565000&$22^4$&253000 \\
    0.9&6.1300&$10 \times 22^3$&725000&$22^4$&543000 \\
    &6.2650&$12 \times 26^3$&504000&$26^4$&256000 \\ \hline

    &6.1250&$8 \times 18^3$&1164000&$22^4$&253000 \\
    1.1&6.2750&$10 \times 22^3$&547000&$22^4$&280000 \\
    &6.4200&$12 \times 26^3$&212000&$26^4$&136000 \\ \hline

    &6.2100&$8 \times 18^3$&1903000&$22^4$&301000 \\
    1.25&6.3600&$10 \times 22^3$&877000&$22^4$&217000 \\
    &6.5050&$12 \times 26^3$&390000&$26^4$&240000 \\ \hline

    &6.3384&$8 \times 18^3$&1868000&$22^4$&544000 \\
    1.5&6.5250&$10 \times 22^3$&1333000&$22^4$&605000 \\
    &6.6500&$12 \times 26^3$&882000&$26^4$&335000 \\ \hline

    &6.5500&$8 \times 18^3$&2173000&$22^4$&534000 \\ 
    2.0&6.7500&$10 \times 22^3$&1671000&$22^4$&971000 \\
    &6.9000&$12 \times 26^3$&1044000&$26^4$&553000 \\ \hline

    &6.9500&$8 \times 26^3$&1300000&$26^4$&433000 \\
    3.0&7.0500&$10 \times 32^3$&563000&$32^4$&148000 \\
    &7.2000&$12 \times 38^3$&317000&$38^4$&60000 \\
\end{tabular}
\label{tb.simulation}
\end{table}

\begin{figure}[t!]
\begin{center}\includegraphics[scale=0.45]{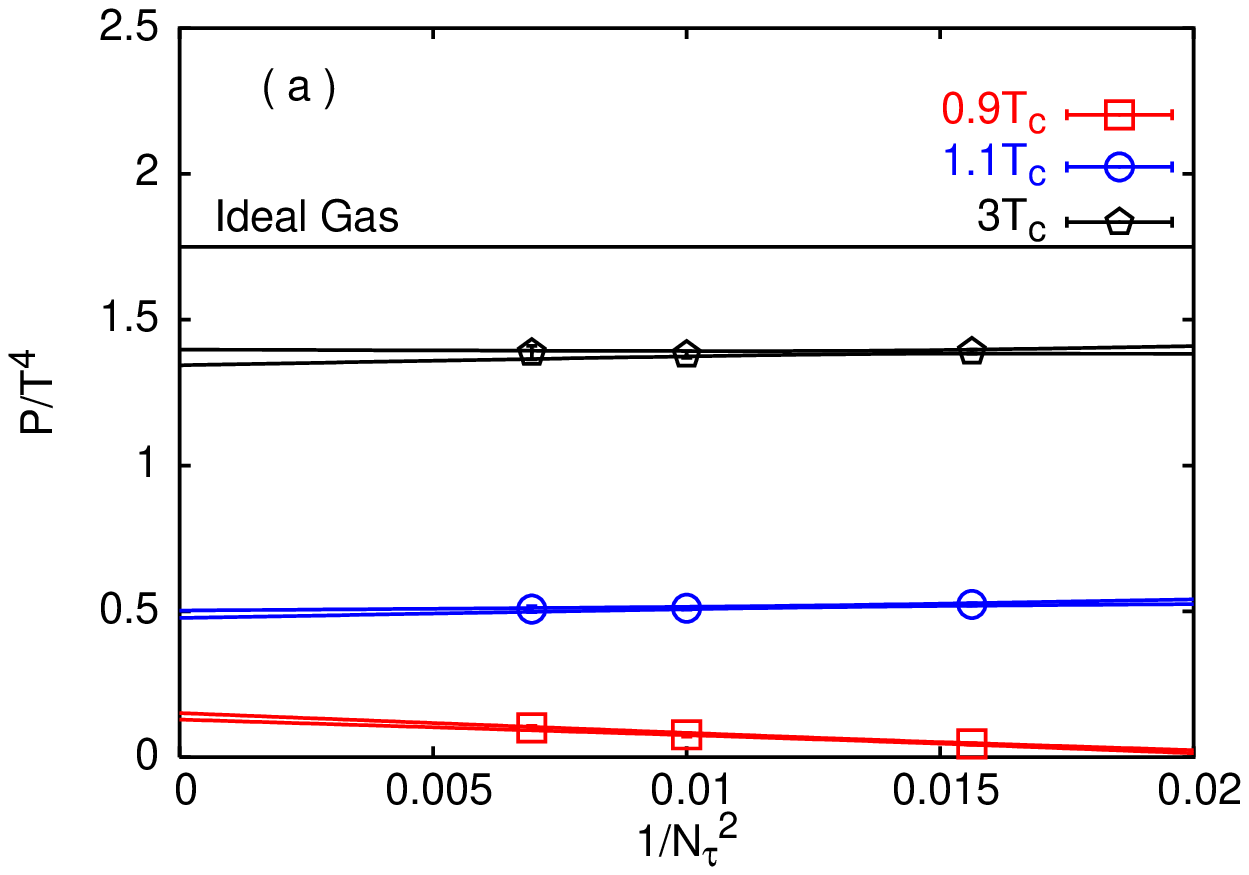}
   \includegraphics[scale=0.45]{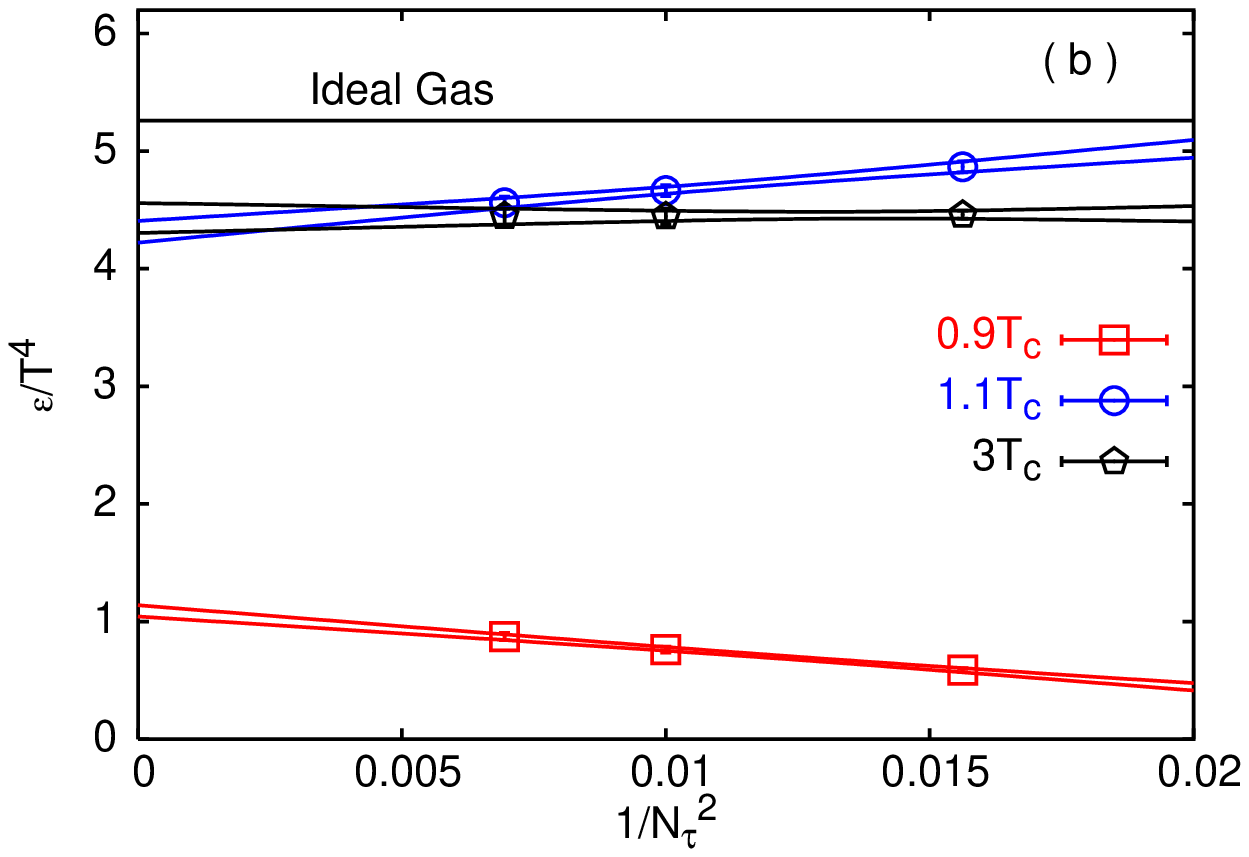}\end{center}
\caption{In the panel (a) we show the dependence of $P/T^4$ on $1/N_\tau^2$
   for different temperature values. In the panel (b) we show the same
   for $\epsilon/T^4$. The 1-$\sigma$ error band of the continuum 
   extrapolations have been indicated by the lines.}
\label{fig.p-e}\end{figure}

We performed $a\to0$ (continuum) extrapolations by linear fits in $a^2 \propto
1/N_{\tau}^2$ at all temperatures using the three values $N_{\tau}=8$,
10, and 12.  In Figure \ref{fig.p-e}(a) we show our data on $P/T^4$ at
finite lattice spacings and the continuum extrapolations for different
temperatures, both above and below $T_c$. We draw attention to the fact
that the pressure is positive on each of the lattices we have used and
also in the $a\to0$ limit. It is an interesting piece of lattice
physics, not relevant to the continuum limit, that the slope of the
continuum extrapolation changes sign at $T_c$. This is also true of the
continuum extrapolation for $\epsilon/T^4$ as shown in Figure
\ref{fig.p-e}(b).  The extrapolation of both $P/T^4$ and $\epsilon/T^4$
between $1.1T_c$ and $3T_c$ are similar to those shown and
have therefore been left out of the figure to avoid clutter.

\begin{figure}[!t]
\begin{center}\includegraphics[scale=0.45]{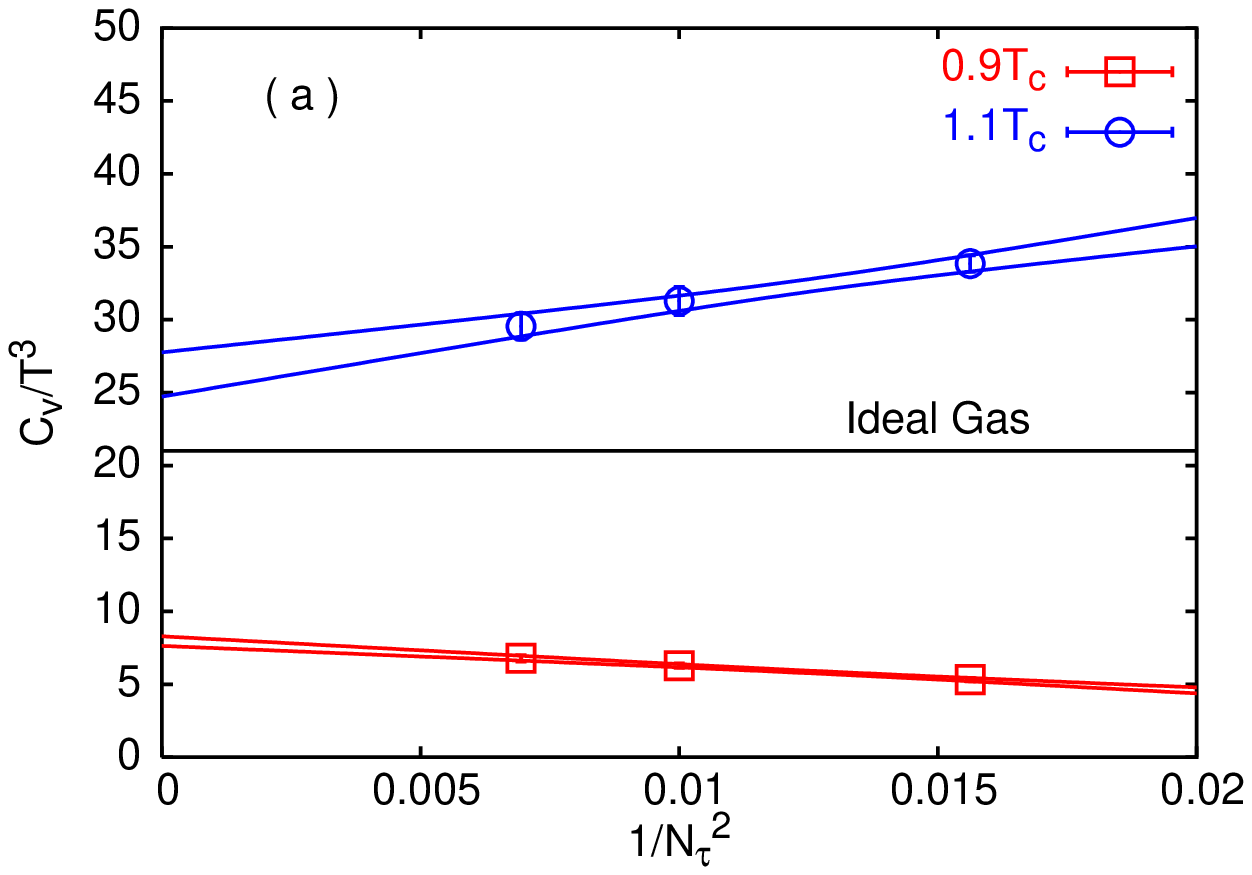}
   \includegraphics[scale=0.45]{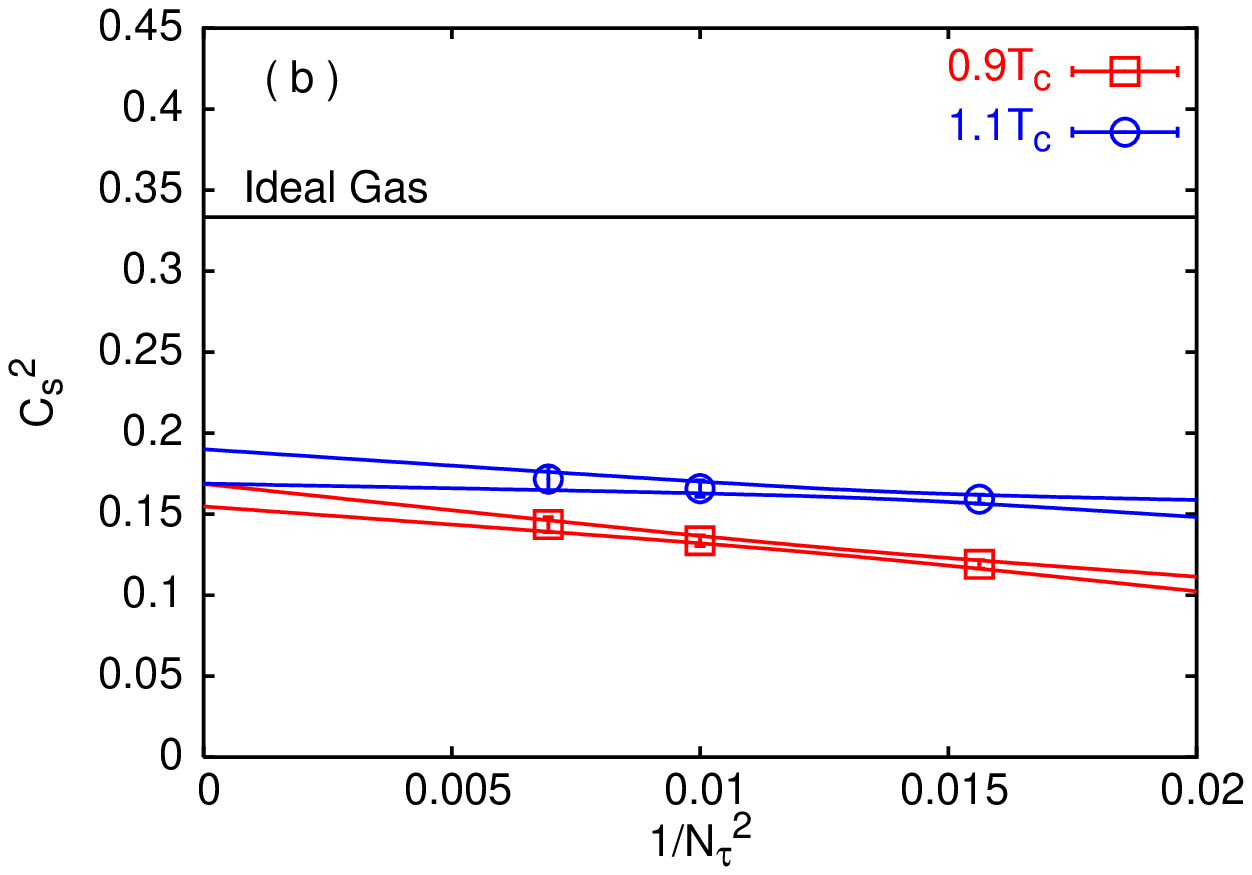}\end{center}
\caption{In the panel (a) we show the dependence of $\cv/T^3$ on $1/N_\tau^2$
   for different values of temperature. In the panel (b) we show the 
   same for $\cs^2$. The 1-$\sigma$ error band of the continuum 
   extrapolations have been indicated by the lines.}
\label{fig.cv-cs}\end{figure}

Similar continuum extrapolations are shown for $\cv/T^3$ and $\cs^2$
in the two panels of Figure \ref{fig.cv-cs}. In all cases, the
continuum extrapolations are smooth, and well fitted by a straight
line in the range of $N_\tau$ used in this study. As mentioned
above, it is interesting lattice physics to see that for $\cv/T^3$
also, the slope of the continuum extrapolation flips sign at $T_c$.
This does not happen for $\cs^2$. Since this is the derivative of
the energy density with respect to the pressure, the slope of this
quantity depends on the slopes of the continuum extrapolation of
$\epsilon/T^4$ and $P/T^4$.

\begin{figure}[!t]
\begin{center}
   \includegraphics[scale=0.45]{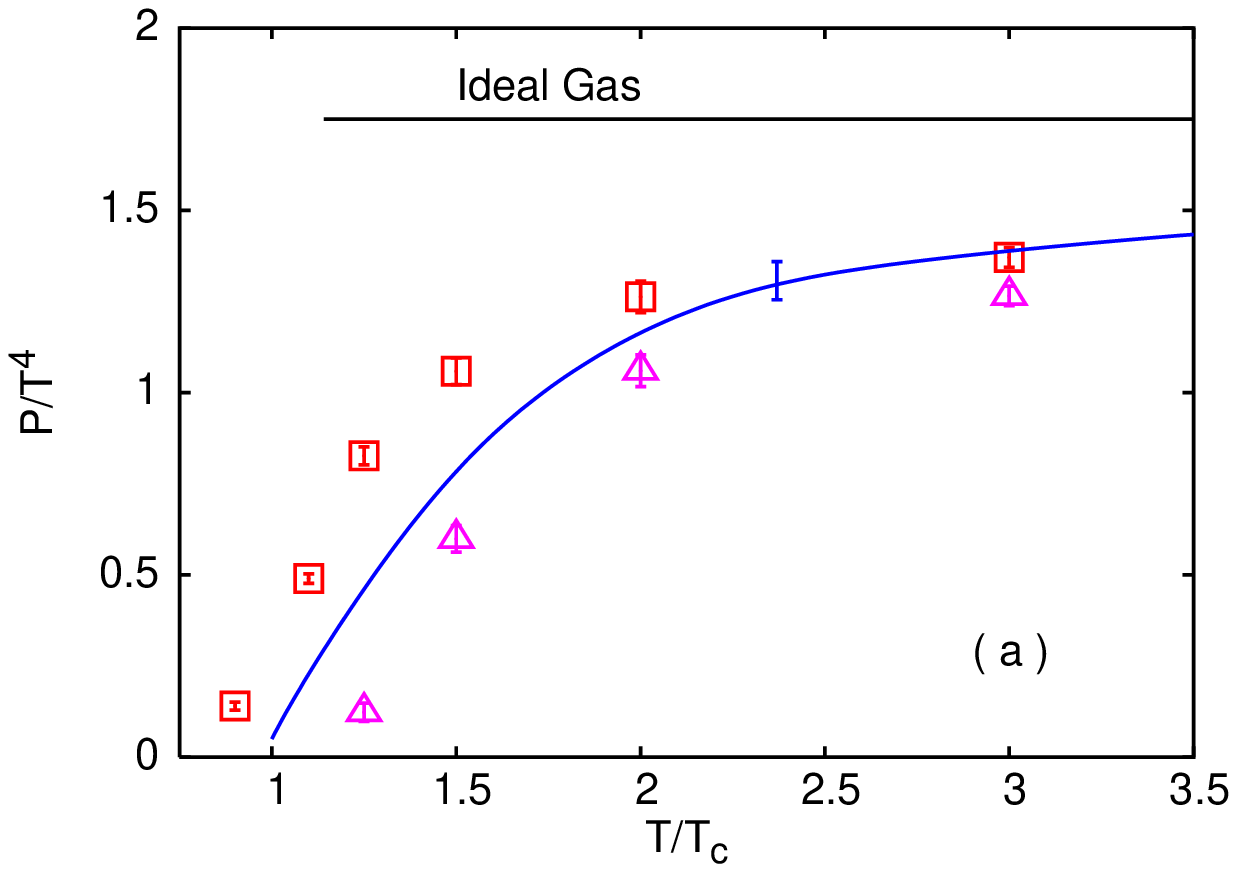}
   \includegraphics[scale=0.45]{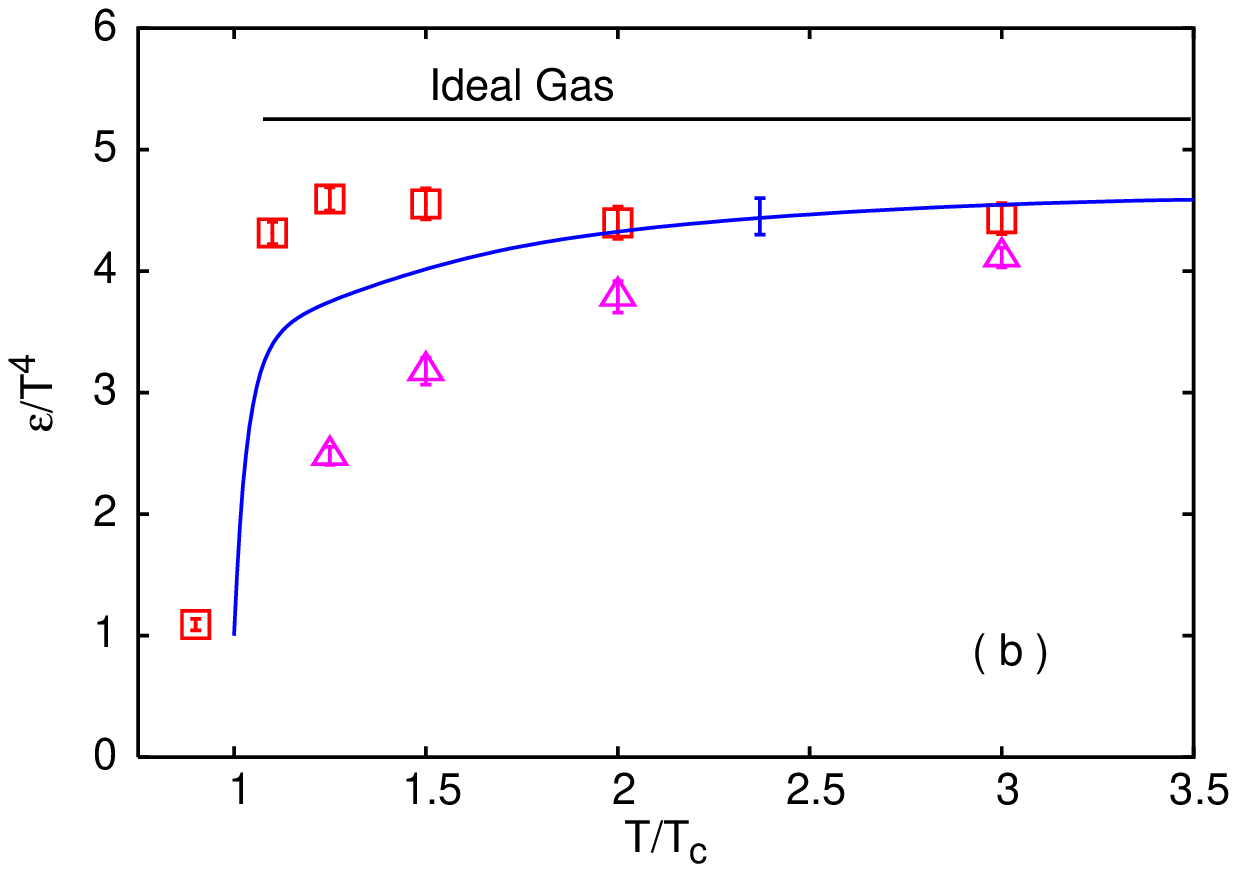} \\
   \includegraphics[scale=0.44]{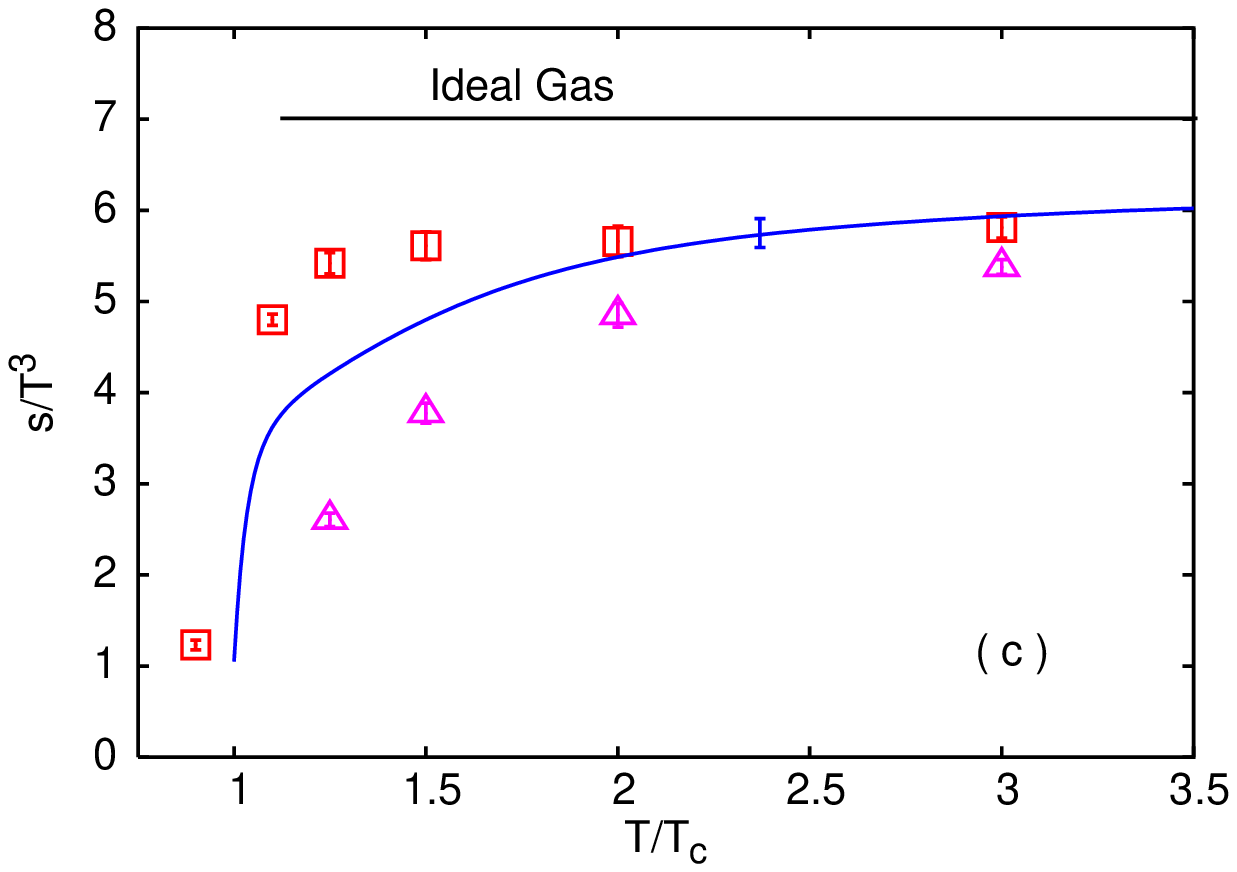}
   \includegraphics[scale=0.45]{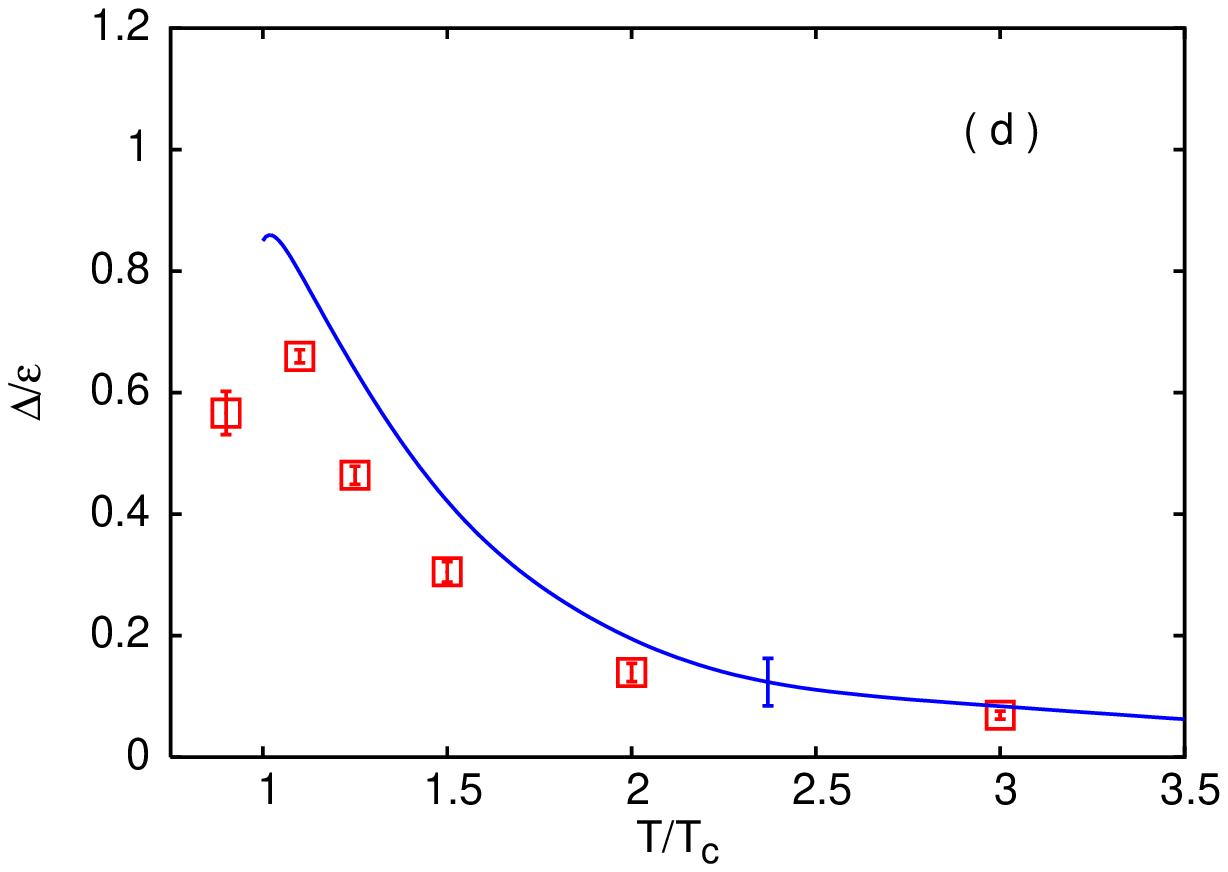} \\
   \includegraphics[scale=0.45]{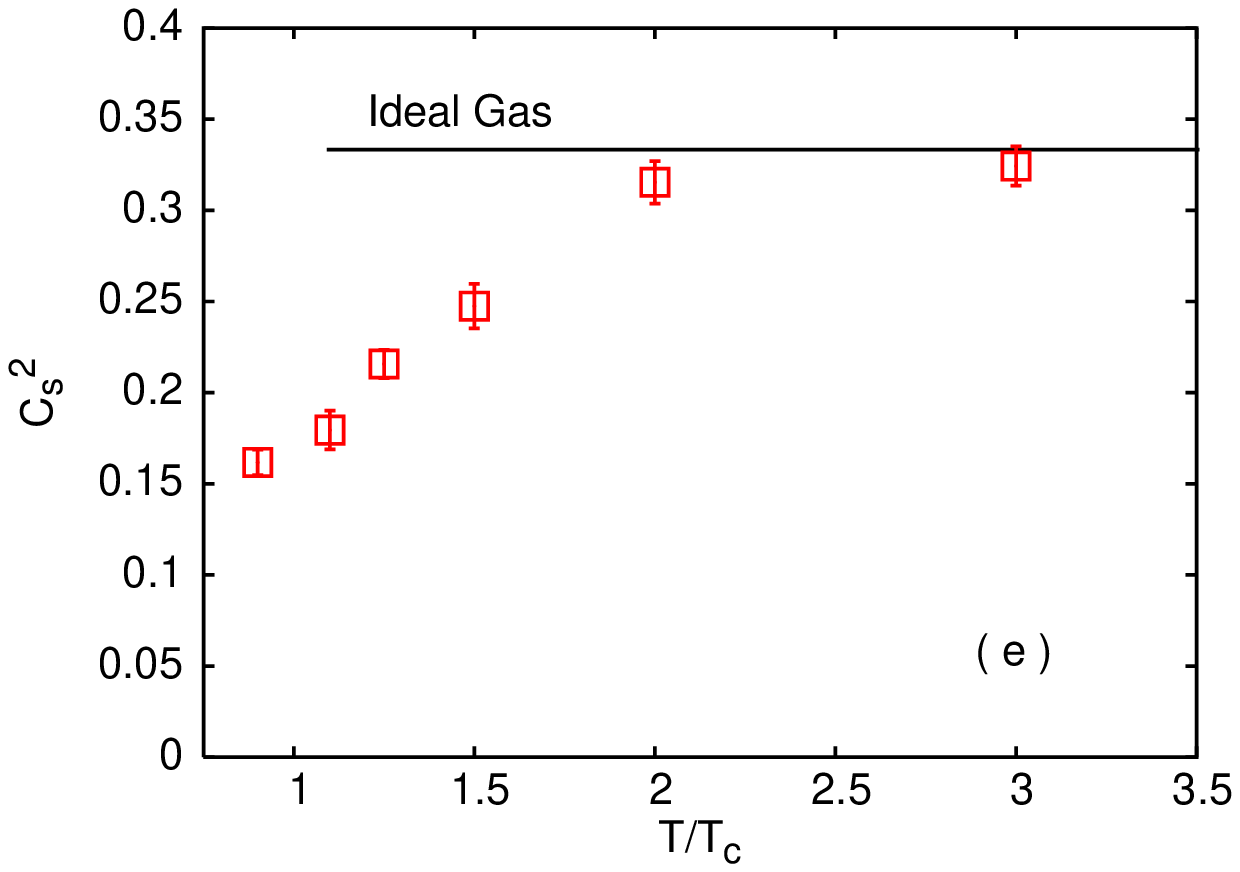}
   \includegraphics[scale=0.45]{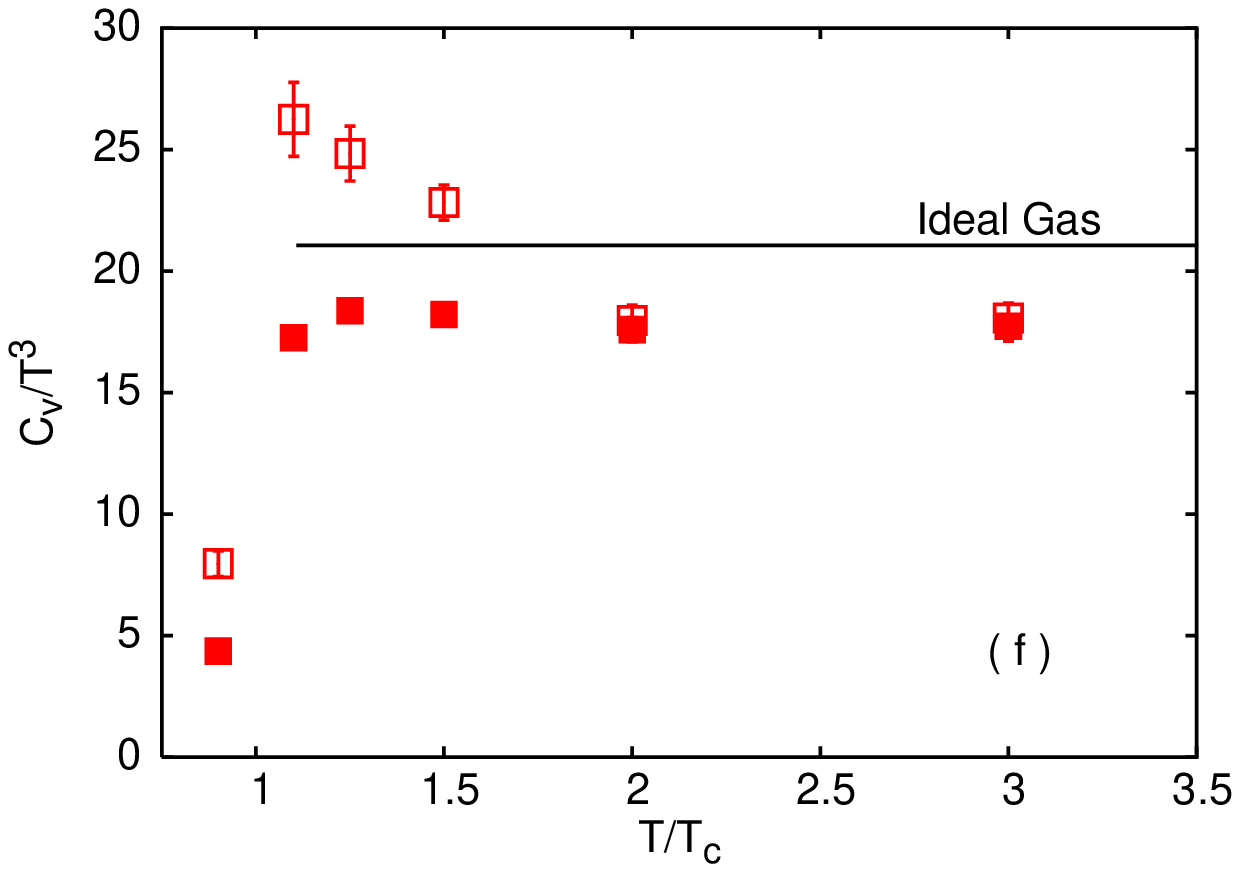}
\end{center}
\caption{We show comparisons between the continuum extrapolated results 
of different
thermodynamic quantities for t-favoured scheme (boxes), the s-favoured scheme
(triangles) and the integral method (line). In panel (d) we show the continuum 
extrapolated values
of the conformal measure $\C$ (boxes). In panel (f) we show a
comparison between our continuum extrapolated results for $\cv/T^3$ (open boxes) 
and that of
$4\epsilon/T^4$ (filled boxes). The data for the integral method has been taken form
Ref.\ [2].}
\label{fig.cont}
\end{figure}

\begin{figure}[!t]
\begin{center}
  \includegraphics[scale=0.45]{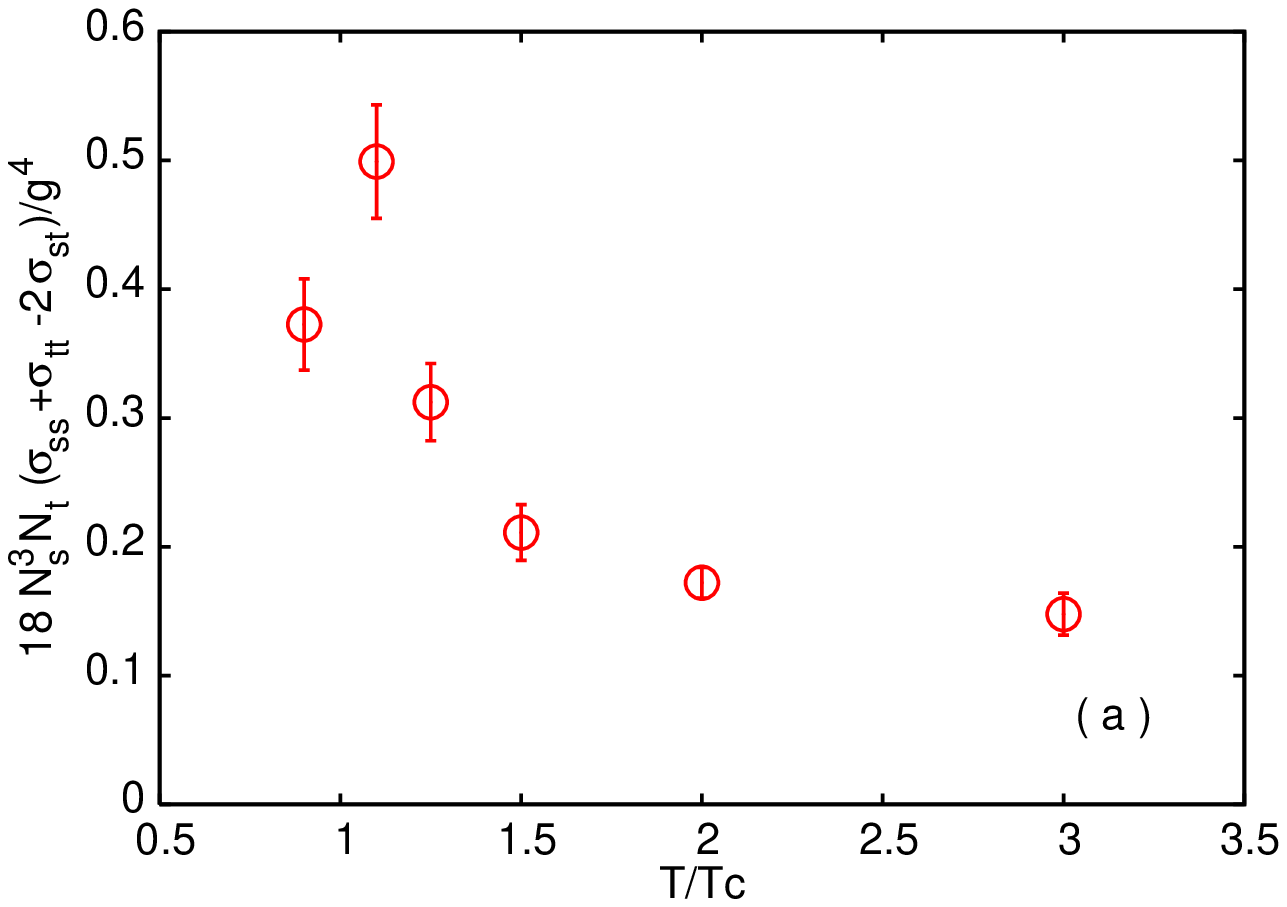}
  \includegraphics[scale=0.48]{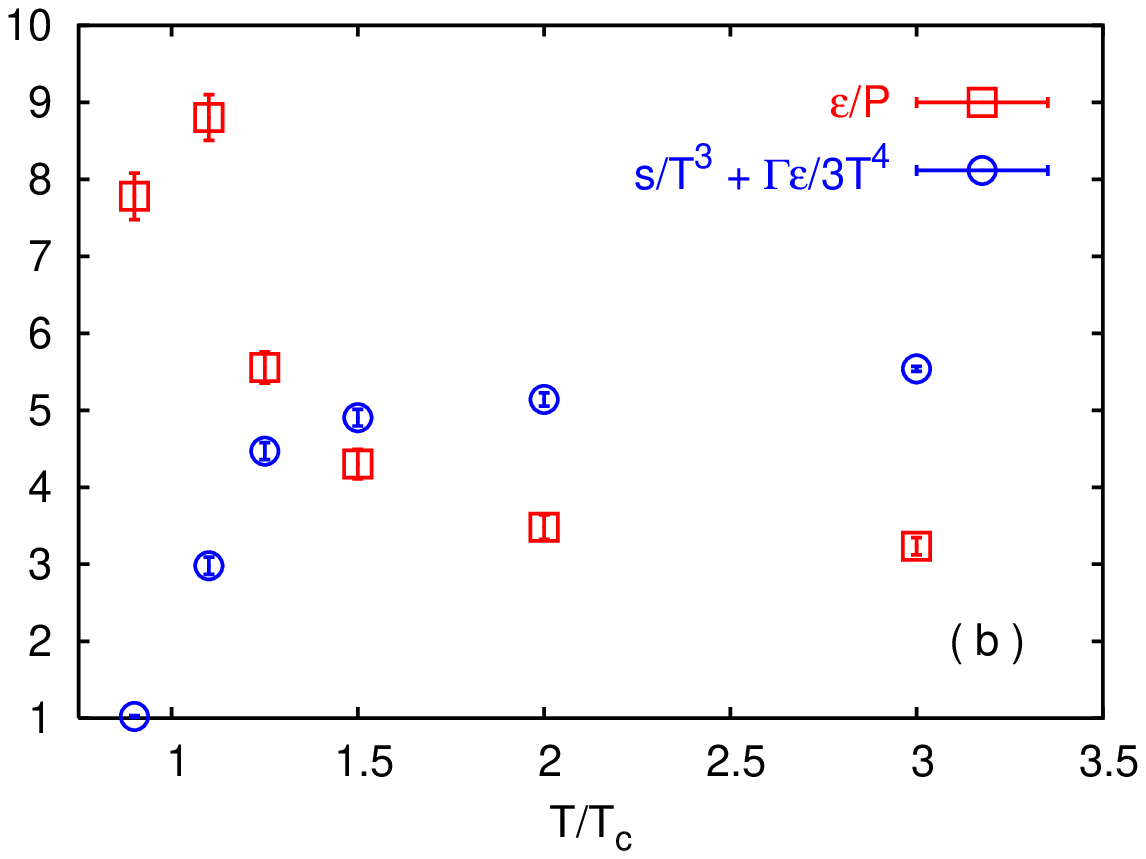}
\end{center}
\caption{In panel (a) we show the temperature dependence of the contribution 
         of one of the covariance terms in $\cv/T^3$. In panel (b) we
         show the the individual contribution of the two factors in Eq.\ 
         (\ref{eq.cvcs}) for $\cv/T^3$. See the text for a detailed discussion.}
\label{fig.var}\end{figure}

The results of continuum extrapolations of our measurements are collected in
Table \ref{tb.cont-values}. It is gratifying to note that the pressure
and the entropy are not only positive in the full temperature range,
but also convex functions of $T$, as required for thermodynamic stability.

In the various panels of Figure \ref{fig.cont} we show a comparison
between the continuum extrapolated results for different quantities obtained using
the t-favoured scheme, s-favoured scheme and the integral method. While
the results of the t-favoured and the s-favoured schemes are obtained
from the analysis of our data, the results of the integral method are
taken form Ref.\ \cite{boyd}.

First we note that unlike the s-favoured differential method, the
t-favoured scheme yields a positive pressure [Figure\
\ref{fig.cont}(a)] at all $T$. There is apparent agreement
between the integral and the t-favoured operator method for $T\ge2T_c$,
both differing from the ideal value by about 20\%. Only at these
temperatures the coupling $\beta$ becomes $\ge6.55$ for all the lattices
(see Table\ \ref{tb.simulation}) that has been used to extract the
continuum extrapolated values in the t-favoured scheme. Hence, from our earlier
discussion it is clear that an agreement between the two methods is
expected to take place at these temperatures. There can be several
causes for the difference between these two methods closer to $T_c$---
(i) The use of one-loop order perturbative Karsch coefficients in the
t-favoured scheme is probably the primary cause for this difference. Use
of larger lattices (\ie larger $\beta$) or inclusion of the effects of
higher order loops in the Karsch coefficients is expected to improve the
agreement. (ii) Another possible source of disagreement is that the
results for the integral method shown here were obtained on coarser
lattices \cite{boyd} than the ones used in this study. (iii) The
integral method assumes that the pressure below some $\beta_0$,
corresponding to some temperature $T<T_c$, is zero. By changing
$\beta_0$ one can change the integral method pressure by a temperature
independent constant. This may restore the agreement close to $T_c$,
although in that case the agreement at the high-$T$ region may get
spoiled. (iv) Also different schemes have been used to define the
renormalized coupling in the two cases. This can also make some
contribution to the different results of the two methods.

Correspondingly, the energy density is harder near $T_c$, showing a
significantly lessened tendency to bend down. This could indicate a
difference in the latent heat determined by the two methods. We shall
return to this quantity in the future. The entropy density is shown in
Figure \ref{fig.cont}(c). Since this is a derived quantity (see Eq.\
\ref{eq.entropy}), it has similar features as those of $P/T^4$ and
$\epsilon/T^4$.

The generation of a scale and the consequent breaking of conformal invariance at
short distances, of the order of $a$, in QCD is, of course, quantified by the
$\beta$-function of QCD.  It has been argued in Ref.\ \cite{swagato}, that the
conformal measure, $\C=\Delta/\epsilon$, parametrizes the departure from the
conformal invariance at the distance scale of order $1/T$. In Figure
\ref{fig.cont}(d) we plot $\cal C$. It is clear that at high temperature, 2--3$T_c$,
conformal invariance is better respected in the finite temperature effective
long-distance theory. Closer to $T_c$ conformal symmetry is badly broken even in the
thermal effective theory. This is consistent with the existence of many mass scales
in the theory as found in Ref.\ \cite{saumen2,hart,forcrand}. It is interesting to note that the
t-favoured scheme yields marginally smaller values of $\cal C$ than the integral
method. Note also the peak in $\cal C$ just above $T_c$; this is the reflection of a
similar peak in $\Delta$.

Figure \ref{fig.cont}(e) shows the continuum extrapolated results
for $\cs^2$.  At temperatures of $2T_c$ and above, the speed of
sound is consistent with the ideal gas value within 95\% confidence
limits.  It is seen that $\cs^2$ decreases dramatically near $T_c$.
Below $T_c$ there is again a rise in $\cs^2$, the numerical values
being very close 10\% below and above $T_c$. In future we plan to
explore in greater detail the region in between.

The behaviour of $\cv/T^3$, shown in Figure \ref{fig.cont}(f), is
the most interesting. At $2T_c$ and above it disagrees strongly
with the ideal gas value, but is quite consistent with the prediction
in conformal theories that $\cv/T^3=4\epsilon/T^4$. Closer to $T_c$,
however, even this simplification vanishes. The specific heat peaks
at $T_c$, consistent with the observation of Refs.\ \cite{saumen,olaf}
that there is a light mode (the thermal scalar, called the $A_1^{++}$)
in the vicinity of $T_c$. Below $T_c$ the specific heat is very
small.

In view of the rise in $\cv/T^3$ near $T_c$, we studied the
contributions of the terms containing different covariances of the
plaquettes. As can be seen from the Eqs. (\ref{eq.dfdt},
\ref{eq.dgdt}), among all the terms containing covariances, the term
$(\varss+\vartt-2\varst)/g^4$ will have the largest contribution to
$\cv/T^3$.  All the other terms containing the covariances are
multiplied either by one of the $c_i'$, or by
$B(\alpha_s)/2\pi\alpha_s^2$ and hence become at least one order of
magnitude smaller than this term.

In Figure \ref{fig.var}(a) we show the contribution of the above term,
as a function of $T$ in the continuum limit. It peaks near
$T_c$,  consistent with the decrease of the $A_1^{++}$ screening mass
mentioned earlier. Since the lattices that we used are significantly
larger than this correlation length, we are in the correct
regime of volumes where the central limit theorem holds for the
fluctuations of the plaquettes. The contribution of this term
is very small: comparable to the errors in $\cv$. The origin of the
peak in $\cv$ therefore lies elsewhere.
In Figure \ref{fig.var}(b) we separately plot the two factors,
$\epsilon/P$ and $s/T^3+\Gamma\epsilon/3T^4$, in the the expression for
$\cv$ in Eq.\ \ref{eq.cvcs}. The factor $s/T^3+\Gamma\epsilon/3T^4$ is
smooth in the whole temperature range, and it is the first factor, $\epsilon/P$,
which has a peak near $T_c$. Rewriting this as $3/(1-{\cal C})$, we can
recognize that the peak in $\cv$ is related to that in $\Delta$.

\section{Discussion} \label{sc.discussion}

\begin{table}[t!]
\caption{Continuum values of some quantities at all temperatures we have explored. The
numbers in brackets are the error on the least significant digit. For the convenience
of the readers here we also list the numerical values of these quantities for an
ideal gas--- $\epsilon/T^4\approx5.26$, $P/T^4\approx1.75$, $s/T^3\approx7.02$,
$\cv/T^3\approx21.06$ and $\cs^2=1/3$. The value of the 't Hooft coupling $g^2N_c$ is
computed at the scale $2\pi T$ using the $T_c/\lambdams$ quoted in Ref.\ [32].}
\begin{tabular}{|c|c||c|c|c||c|c|} \hline
   $T/T_c$&$g^2 N_c$&$\epsilon/T^4$&$P/T^4$&$s/T^3$&$\cv/T^3$&$\cs^2$\\ \hline
   0.9&11.5(3)&1.09(4)&0.14(1)&1.23(5)&8.0(5)&0.162(7) \\
   1.1&10.4(2)&4.31(9)&0.49(1)&4.80(6)&26(2)&0.18(1) \\
   1.25&9.8(2)&4.6(1)&0.82(2)&5.4(1)&25(1)&0.21(1) \\
   1.5&9.0(1)&4.5(1)&1.06(4)&5.6(2)&22.8(7)&0.25(1) \\
   2.0&8.1(1)&4.4(1)&1.26(4)&5.7(2)&17.9(7)&0.31(1) \\
   3.0&7.0(1)&4.4(1)&1.37(3)&5.8(1)&17.9(8)&0.32(1) \\
\end{tabular}
\label{tb.cont-values}
\end{table}

In this paper we have proposed a modification, \viz the t-favoured scheme, of
the differential method for the computation of the QCD equation of state. We
have shown that this improvement gives positive pressure for all temperatures
and $N_{\tau}$ used, even when the older s-favoured differential method
\cite{engels} gives negative pressure. Note that this is so in spite of the use
of the same one-loop order perturbative values for the couplings in both cases.
Using the t-favoured differential method and by extrapolating to the $a\to0$ 
(continuum) limit we obtain the energy density and pressure for a pure gluonic 
theory in the temperature range $0.9 \le T/T_c \le 3$. 
These differ from their respective ideal gas values by
about 20\% at $3T_c$, and by much more as one approaches $T_c$. On comparing
our results with those of the integral method \cite{boyd}, we found that ours
are larger for $T<2T_c$. The primary reason behind this disagreement seems to
be our use of perturbative couplings. Hence the agreement between the
t-favoured scheme and the integral method is expected to improve by going to
larger temporal lattice sizes or equivalently to smaller lattice spacings. 

We have also extended the t-favoured scheme to compute the continuum 
extrapolated results
of the specific heat at constant volume and the speed of sound. We
found that $\cv$ peaks near $T_c$ where, in addition, $\cs$ becomes
small.  Our results are collected together in Table \ref{tb.cont-values}
and Figure \ref{fig.cont}. The most robust quantity on the equation
of state in all lattice computations is $\Delta$, and the most
interesting (and also stable) feature seen to date is the peak in
$\Delta$ just above $T_c$. Apart from influencing the EOS, it
manifests itself as a peak in $\cv$. Since $\cv$ could be directly
measurable through energy or effective temperature fluctuations in
heavy-ion collisions, understanding $\Delta$ should be one of the
prime goals of theory. Unfortunately, it seems that at present no
tool other than lattice computations are available for this task.
Even models of this important and stable phenomenon are lacking.

\begin{figure}[!t]
\begin{center}
  \includegraphics[scale=0.45]{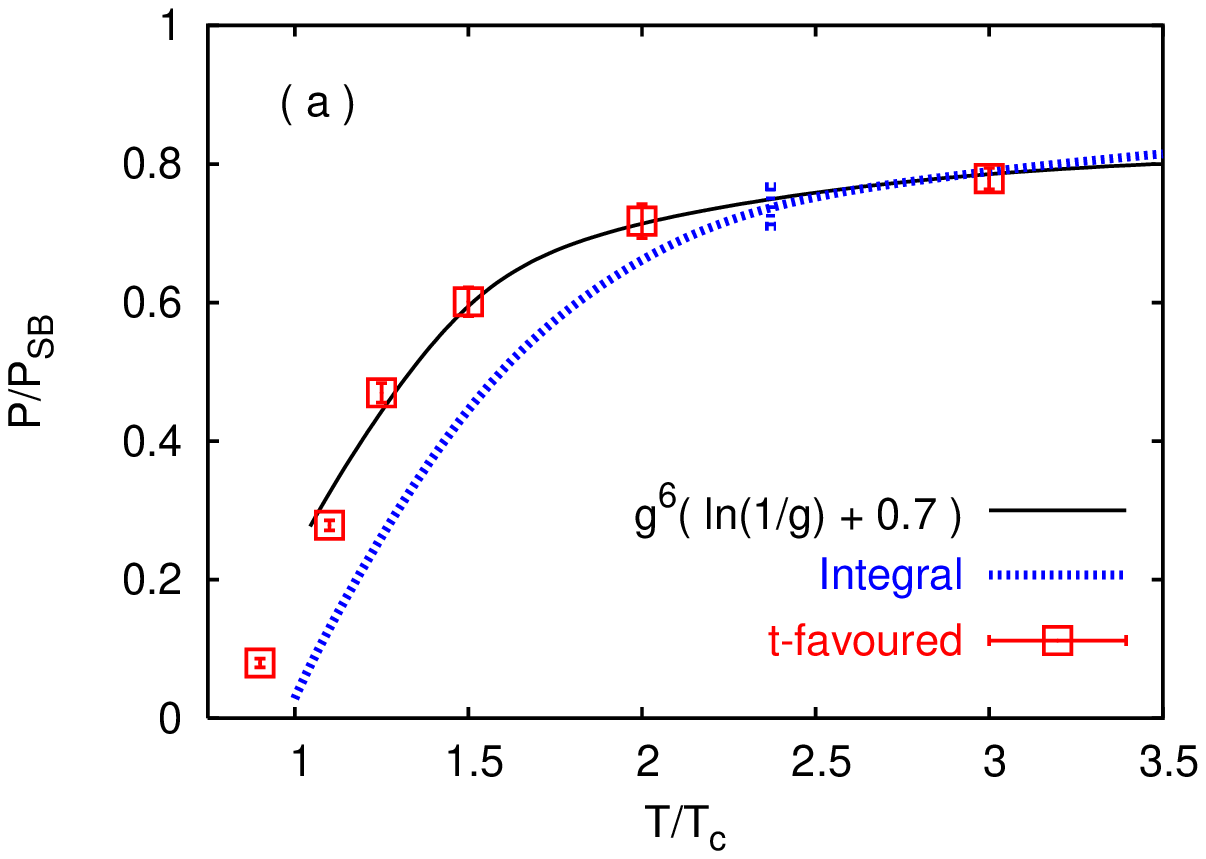}
  \includegraphics[scale=0.45]{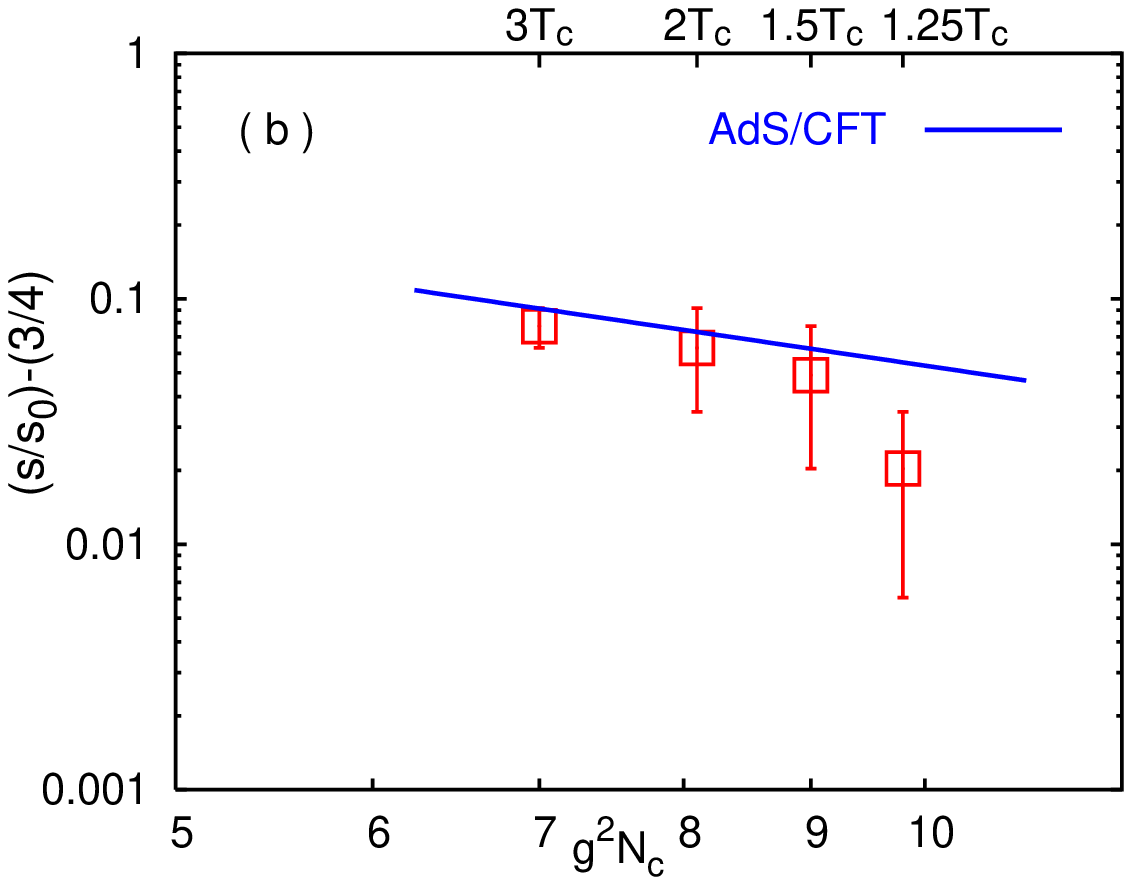}
\end{center}
\caption{In panel (a) we compare the pressures obtained by t-favoured method (boxes),
integral method (dotted line) and the $g^6\ln(1/g)$ order perturbative expansion
(solid line) of The data for the integral method and the perturbative expansion are
taken from Ref.\ [2] and Ref.\ [40] respectively. The values of
the $T/\lambdams$ in Ref.\ [40] has been converted to $T/T_c$ using the
$T_c/\lambdams$ quoted in Ref.\ [32].  In panel (b) we show the deviation of
$s/s_0$ from $3/4$ (boxes) as a function of the 't Hooft coupling. We also show the
prediction of Eq.\ (1) (solid line).}
\label{fig.pert-cft}
\end{figure}

In view of the fact the perturbation theory fails to reproduce the lattice data on
EOS, specially close to $T_c$, it is interesting to compare our t-favoured scheme
results with that of the perturbation theory.  In Figure \ref{fig.pert-cft}(a) we
compare the pressure obtained in the t-favoured method with that from a dimensionally
reduced theory, matched with the 4-d theory perturbatively up to order
$g^6\ln(1/g)$ \cite{kajantie}. Writing $P_{SB}$ for the ideal gas (Stefan-Boltzmann)
value of the pressure, the ratio for $P/P_{SB}$ found in the dimensionally reduced
theory \cite{kajantie} has an undetermined adjustable constant, $c$. The pressure
determined through dimensional reduction agrees with our results almost all the way
down to $T_c$, for that value of the constant ($c=0.7$) for which it matches with the
integral method in the high temperature range. In future it would be interesting to
check whether an equally good description is available in this approach for the full
entropy. This would be a non-trivial extension because perturbation theory misses
$\Delta$ completely. The question, therefore, addresses the non-perturbative dynamics
of the dimensionally reduced theory.

\begin{figure}[t!]
\begin{center}
\includegraphics[scale=0.7]{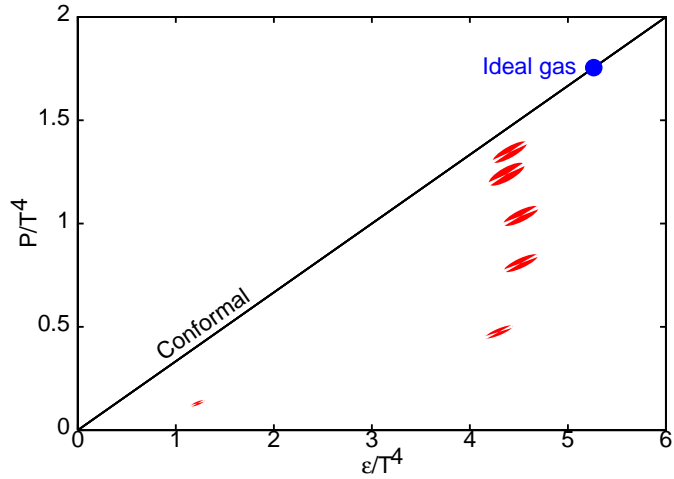}
\end{center}
\caption{The equation of state of QCD matter. The diagonal line denotes possible EOS
for theories with conformal symmetry. The circle on the diagonal denotes the ideal
gluon gas, whose EOS in this form is temperature independent. The ellipses denote
66\% error bounds on the measured EOS (see Ref.\ [41]). The ratio of the axes is a
measure of the covariance in the measurements of $\epsilon/T^4$ and $P/T^4$. The
wedges piercing these ellipses have average slope $\cs^2$, and the opening half-angle
of these wedges indicate the error in $\cs^2$.}
\label{fig.eos}
\end{figure}

The strong coupling result in Eq. (\ref{eq.sym}) of Ref.\ \cite{gubser} can be 
compared with our data on the entropy density, $s/T^3$. This has to be 
done in an appropriate window of $T$ where the 't Hooft
coupling $g^2N_c$ is large and $\cal C$ is small. The strong coupling
series is an expansion in $(g^2N_c)^{-1/2}$. For $N=4$ SYM, the
first term vanishes due to a delicate cancellation and the series
starts with the $(g^2N_c)^{-3/2}$ term \cite{gubser}.  When some
of the supersymmetry is broken, this cancellation need not occur
and the series could start with a term in $(g^2N_c)^{-1/2}$. Needless
to say, the theory we are studying here, pure QCD, lacks 
supersymmetry.  In Figure \ref{fig.pert-cft}(b) we show the deviation
of $s/s_0$ from $3/4$ as a function of the 't Hooft coupling ($s$ and
$g^2N_c$ are listed in Table \ref{tb.cont-values}). Also shown is the
prediction of Eq. (\ref{eq.sym}). Comparison of our data with the latter 
shows that the AdS/CFT based theory agrees with our data for 
$g^2N_c < 9$, or in other words for $\C < 0.3$.
As a partial summary of our results, we show the equation of state
in Figure \ref{fig.eos} in the form of a plot of
$P/T^4$ against $\epsilon/T^4$, useful for hydrodynamics.
In this plot, the ideal gas for fixed number of colours is represented
by a single point which is independent of $T$, and theories with
conformal symmetry by the line $\epsilon=3P$. Pure gauge QCD lies
close to the conformal line at high temperature, as shown, but
deviates strongly nearer $T_c$.

The slope of the wedges piercing the ellipses indicates the speed
of sound--- when these are parallel to the conformal line then
$\cs^2=1/3$. This is clearly the case at high temperature. However,
there is an increasing flattening of the axis, denoting a drop in
$\cs^2$ as one approaches $T_c$. Note that the slope of the curve
joining the middle points of the ellipses does not give $\cs^2$,
since the plot is of $\epsilon/T^4$ against $P/T^4$. In a plot of
$\epsilon$ against $P$, it would have been correct to assume that
the slope gives $\cs^2$.

Two other physically important effects can be read off the figure.
First, the softening of the equation of state just above $T_c$ is shown
by the rapid drop in pressure at roughly constant $\epsilon/T^4$.
Second, a large latent heat is indicated by the jump between the last
two points, at almost the same pressure but very different energy
densities. 

A final piece of physics can be deduced from the fact that the low
temperature phase shows a very small $P/T^4$ at a significantly
large value of $\epsilon/T^4>1$ just below $T_c$. This is an
indication that there are very massive modes in the hadron gas which
contribute large amounts to $\epsilon$ without contributing to $P$.
The small value of $\cv/T^3$ at the same $T$ also indicates that
the energy required to excite the next state is rather large. We
have mentioned already that the observations just above $T_c$ are
compatible with the known spectrum of excitations in pure gauge QCD
\cite{saumen2}.

\acknowledgments
We would like to thank Frithjof Karsch for helpful discussions.

\appendix
\section{Discussion on the Karsch coefficients} \label{app.a}

The Karsch coefficients ($c_i$) are differences between the anisotropic and
isotropic lattice couplings and hence do not depend on the scale $a$ of the
isotropic lattice, but only on the the anisotropic parameter $\xi$. One can see
this directly from the derivations in Ref.\ \cite{karsch}, where these have
been evaluated up to one-loop order in the perturbation theory. For any
arbitrary $\xi\ne1$, all integrals contributing in the effective action
$S_{eff}(a,\xi)$, mentioned in Eq.\ (2.22) of Ref.\ \cite{karsch}, are
independent of the scale the $a$. The dependence of $a$ are only encoded
implicitly in the couplings $g_i^{-2}(a,\xi)$. Hence $S_{eff}(a,\xi)$ of Eq.\
(2.22) of Ref.\ \cite{karsch} is equally valid for $a=a_\tau$. The values of
the Karsch coefficients have been evaluated by imposing the condition $\Delta
S_{eff}=S_{eff}(a,\xi)-S_{eff}(a,1)=0$, which is again independent of the scale
$a$. Hence the one-loop order Karsch coefficients for both the case $a=a_s$
($s-$favoured scheme) and $a=a_\tau$ ($t-$favoured scheme) are the same. 

Nevertheless, we derive this equality explicitly in the following. Let us
assume that the one-loop order perturbative expansions for $g_i^2$'s, around
the isotropic lattice coupling $g$, have the following forms
\beqa
  g_i^{-2}(a_s,\xi) &=& g^{-2}(a_s) + c_i(\xi) + O[g^2(a_s)] ,
  \qquad {\rm and} \nonumber \\
  g_i^{-2}(a_\tau,\xi) &=&  g^{-2}(a_\tau) + \alpha_i(\xi) +
  O[g^2(a_\tau)] .
\label{eq.ap1}
\eeqa

Our claim is that $\left[\partial c_i(\xi)/\partial
\xi\right]_{a_s}=\left[\partial \alpha_i(\xi)/\partial
\xi\right]_{a_\tau}$. In order to prove it we make a Taylor series
expansion of $g_i(a_s,\xi)$ around $a_s=a_\tau$, at any fixed $\xi\ne1$
\beq
  g_i^{-2}(a_s,\xi) = g_i^{-2}(a_\tau,\xi) +
  \sum_{n=1}^{\infty} \frac{(a_s-a_\tau)^n}{n!} \left[\left.\frac{\partial^n
  g_i^{-2}(x,\xi)}{\partial x^n}\right|_\xi\right] _{x=a_\tau} . 
\label{eq.ap2}
\eeq
A $\xi$ derivative at constant $a_s$, on Eq. (\ref{eq.ap2}) yields
\beqa
  \left. \frac{\partial g_i^{-2}(a_s,\xi)}{\partial \xi} \right|_{a_s} &=&
  \left. \frac{\partial g_i^{-2}(a_\tau,\xi)}{\partial \xi}
  \right|_{a_s} + \sum_{n=1}^{\infty} \frac{na_s^n}{n!\xi^2}
  \left(1-\frac{1}{\xi}\right)^{n-1} \left.\frac{\partial^n
  g_i^{-2}(a_\tau,\xi)}{\partial a_\tau^n}\right|_\xi \nonumber \\
  &+&\sum_{n=1}^{\infty} \frac{a_s^n}{n!}\left(1-\frac{1}{\xi}\right)^n
  \frac{\partial}{\partial \xi}\left[ \left.\frac{\partial^n
  g_i^{-2}(a_\tau,\xi)}{\partial a_\tau^n}\right|_\xi \right]_{a_s} .
\label{eq.ap3}
\eeqa
While $\left[\partial g(a_s)/\partial
\xi\right]_{a_s} =0$, $\left[\partial g(a_\tau)/\partial
\xi\right]_{a_s}=\left[\partial g(a_s/\xi)/\partial
\xi\right]_{a_s}\ne0$, from Eq.\ (\ref{eq.ap1}) it follows that 
\beqa
  \left.\frac{\partial g_i^{-2}(a_\tau,\xi)}{\partial \xi}\right|_{a_s}
  &=& \left.\frac{\partial g^{-2}(a_\tau)}{\partial \xi}\right|_{a_s}
  + \left.\frac{\partial \alpha_i(\xi)}{\partial \xi}\right|_{a_s}
  \nonumber \\ &=& \frac{\partial}{\partial \xi}\left[ g^{-2}(a_s) +
  \sum_{n=1}^{\infty}\frac{(a_\tau-a_s)^n}{n!} \frac{\partial^n
  g^{-2}(a_s)}{\partial a_s^n} \right]_{a_s} + \left.\frac{\partial
  \alpha_i(\xi)}{\partial \xi}\right|_{a_s} \nonumber \\ &=&
  -\sum_{n=1}^{\infty}\frac{na_s^n}{n!\xi^2}\left(\frac{1}{\xi}-1\right)^{n-1}
  \frac{\partial^n g^{-2}(a_s)}{\partial a_s^n} + \left.\frac{\partial
  \alpha_i(\xi)}{\partial \xi}\right|_{a_s} .
\label{eq.ap4}
\eeqa
Substituting Eq.\ (\ref{eq.ap4}) in Eq.\ (\ref{eq.ap3}) and using
relations in Eq.\ (\ref{eq.ap1}) to calculate the various
derivatives one obtains
\beqa   
  \left. \frac{\partial c_i(\xi)}{\partial \xi} \right|_{a_s} &=&
  -\sum_{n=1}^{\infty}\frac{na_s^n}{n!\xi^2}\left(\frac{1}{\xi}-1\right)^{n-1}
  \frac{\partial^n g^{-2}(a_s)}{\partial a_s^n} + \left.\frac{\partial
  \alpha_i(\xi)}{\partial \xi}\right|_{a_s} \nonumber \\
  &+& \sum_{n=1}^{\infty}
  \frac{na_s^n}{n!\xi^2} \left(1-\frac{1}{\xi}\right)^{n-1}
  \left.\frac{\partial^n g^{-2}(a_\tau)}{\partial a_\tau^n}\right|_\xi
  \nonumber \\ &+& \sum_{n=1}^{\infty} \frac{a_s^n}{n!}\left(1-\frac{1}{\xi}\right)^n
  \frac{\partial}{\partial \xi}\left[\frac{\partial^n
  g^{-2}(a_\tau)}{\partial a_\tau^n}\right]_{a_s} .
\eeqa
Finally, taking the $\xi\to1$ limit, {\it i.e.} setting $a_s=a_\tau$, one has
\beq
  \left. \frac{\partial c_i(\xi)}{\partial \xi} \right|_{a_s} = 
   \left.\frac{\partial \alpha_i(\xi)}{\partial \xi}\right|_{a_s}
\label{eq.ap5}
\eeq
A variable transformation from $\{a_s,\xi\}$ to $\{a_\tau,\xi\}$
gives $\xi\left(\partial/\partial \xi\right)_{a_s}\equiv
\xi\left(\partial/\partial \xi\right)_{a_\tau} -
a_\tau\left(\partial/\partial a_\tau\right)_\xi$. Using it on Eq.\
(\ref{eq.ap5}) one conclusively proves that
\beq
  \left. \frac{\partial c_i(\xi)}{\partial \xi} \right|_{a_s} = 
  \left.\frac{\partial \alpha_i(\xi)}{\partial \xi}\right|_{a_\tau}
\eeq


\begin{thebibliography}{99}

\bibitem{rhic}
K.\ Adcox \etal [PHENIX Collaboration], \np A {\bf 757}, 184 (2005); \\
I.\ Arsene \etal [BRAHMS collaboration], \np A{\bf 757}, 1 (2005); \\
B.\ B.\ Back \etal [PHOBOS collaboration], \np A {\bf 757}, 28 (2005); \\ 
J.\ Adams \etal [STAR Collaboration], \np A {\bf 757}, 102 (2005).
\bibitem{boyd}
   G.\ Boyd \etal, \prl {\bf 75}, 4169 (1995); \np B {\bf 469}, 419 (1996).
\bibitem{swagato}
   R.\ V.\ Gavai, S.\ Gupta and S.\ Mukherjee, \pr D {\bf 71}, 074013 (2005).
\bibitem{gupta1}
   S.\ Gupta, \pl B {\bf 597}, 57 (2004).
\bibitem{iitk}
   S.\ Gupta, {\sl Pramana\/} {\bf 61}, 877 (2003).
\bibitem{pert}
   P.\ Arnold and C.\ Zhai, \pr D {\bf 50}, 7603 (1994); {\sl ibid.\ } 
    D {\bf 51}, 1906 (1995);\\
   E.\ Braaten and A.\ Nieto, \pr D {\bf 53}, 3421 (1996);\\
   K.\ Kajantie \etal, \prl {\bf 86}, 10 (2001); \jhep {\bf 0304}, 036 (2003).
\bibitem{peshier}
   A.\ Peshier, \np A {\bf 702}, 128 (2002).
\bibitem{bir}
   J.\ O.\ Andersen \etal, \pr D {\bf 66}, 085016 (2002);\\
   J.-P.\ Blaizot, E.\ Iancu and A.\ Rebhan, \pr D {\bf 63}, 065003(2001); \\
   K.\ Kajanie \etal, \prl {\bf 86}, 10 (2001).
\bibitem{pisarski}
   A.\ Dumitru, J.\ Lenaghan and R.\ D.\ Pisarski, \pr D {\bf 71}, 074004 (2005).
\bibitem{grossman}
   B.\ Grossman \etal, \np B {\bf 417}, 289 (1994).
\bibitem{kari}
   K.\ Kajantie \etal, \prl {\bf 79}, 3130 (1997).
\bibitem{saumen}
   S.\ Datta and S.\ Gupta, \np B {\bf 534}, 392 (1998).
\bibitem{owe}
   M.\ Laine and O.\ Philipsen, \pl B {\bf 459}, 259 (1999).
\bibitem{saumen2}
   S.\ Datta and S.\ Gupta, \pr D {\bf 67}, 054503 (2003).
\bibitem{hart}
   A.\ Hart, M.\ Laine and O.\ Philipsen, \np B {\bf 586}, 443 (2000).
\bibitem{forcrand}
  P.\ de Forcrand \etal [QCD-TARO Collaboration], \pr D {\bf 63}, 054501 (2001).
\bibitem{gubser}
   S.\ S.\ Gubser, I.\ R.\ Klebanov and A.\ A.\ Tseytlin, \np B {\bf 534},
   202 (1998); \\ I.\ R.\ Klebanov, hep-th/0009139.
\bibitem{son}
   P.\ Kovtun, D.\  T.\ Son and A.\ Starinets, \jhep {\bf 0310}, 064 (2003).
\bibitem{teaney}
   D.\ Teaney, \pr C {\bf 68}, 034913 (2003).
\bibitem{nakamura}
   A.\ Nakamura and S.\ Sakai, \prl {\bf 94}, 072305 (2005).
\bibitem{meyer}
H.\ B.\ Meyer, \pr D {\bf 76}, 101701 (2007), {\sl ibid.\ } arXiv:0710.3717
[hep-lat].
\bibitem{engels}
   J.\ Engels \etal, \np B {\bf 205 [FS5]}, 545 (1982).
\bibitem{deng}
   Y.\ Deng, \np B ({\sl Proc.\ Suppl.\/}) {\bf 9}, 334 (1989).
\bibitem{engels1}
  J.\ Engels, J.\ Fingberg, F.\ Karsch, D.\ Miller and M.\ Weber,
  \np B {\bf 252}, 625 (1990).
\bibitem{numrec}
   W.\ H.\ Press \etal, {\sl Numerical Recipes\ } (Cambridge University
   Press, Cambridge, 1986.)
\bibitem{ejiri}
   S.\ Ejiri, Y.\ Iwasaki and K.\ Kanaya, \pr D {\bf 58}, 094505 (1998).
\bibitem{stodolsky}
   L.\ Stodolsky, \prl {\bf 75}, 1044 (1995).
\bibitem{blaizot}
   R.\ S.\ Bhalerao, J.\ P.\ Blaizot, N.\ Borghini and J.\ Y.\
   Ollitrault, \pl B {\bf 627}, 49 (2005).
\bibitem{landau}
   See, for example, L.\ D.\ Landau and E.\ M.\ Lifschitz, {\sl Fluid
   Mechanics\ }, (Reed Elsevier
   plc, Oxford, 1998), p. 254.
\bibitem{gock}
   R.\ V.\ Gavai and A.\ Gocksch, \pr D {\bf 33}, 614 (1986).
\bibitem{aniso}
   See, for example, S.\ Aoki \etal, \np {\sl (Proc.\ Suppl.)\ } B {\bf 106},
   477 (2002).
\bibitem{hasenfratz}
   A.\ Hasenfratz and P.\ Hasenfratz, \np B {\bf 193}, 210 (1981).
\bibitem{karsch}
   F.\ Karsch, \np B {\bf 205 [FS5]}, 285 (1982).
\bibitem{gavai}
   S.\ Datta and R.\ V.\ Gavai, \pr D {\bf 60}, 034505 (1999). 
\bibitem{gupta}
   S.\ Gupta, \pr D {\bf 64},  034507 (2001).
\bibitem{lepage}
   G.\ P.\ Lepage and P.\ B.\ Mackenzie, \pr D {\bf 48},  2250 (1993).
\bibitem{edwards}
   R.\ G.\ Edwards, U.\ M.\ Heller and T.\ R.\ Klassen, \np B {\bf 517},
   377 (1998).
\bibitem{heller}
   U.\ M.\ Heller and F.\ Karsch, \np B {\bf 251 [FS13]}, 254 (1985).
\bibitem{ack}
   We thank Kari Rummukainen, Keijo Kajantie and J\"urgen Engels for
   discussions on this point.
\bibitem{engels2}
J.\ Engels, F.\ Karsch and T.\ Scheideler, \np B {\bf 564}, 303 (2000).   
\bibitem{saumen1}
   S.\ Datta and S.\ Gupta, \pl B {\bf 471}, 382 (2000).
\bibitem{olaf}
   O.\ Kaczmarek \etal, \pr D {\bf 62}, 034021 (2001).
\bibitem{kajantie}
   K.\ Kajantie \etal, \pr D {\bf 67}, 105008 (2003).
\bibitem{lyon}
   L.\ Lyons, {\sl Statistics for nuclear and particle physicists\ },
   (Cambridge University Press, Cambridge, 1992.)

\end{thebibliography}
\end{document}